\documentclass[11pt,authoryear]{elsarticle}

\usepackage{tocloft}
\usepackage[toc,page]{appendix}
\usepackage{enumerate}
\usepackage[ntheorem]{empheq} 
\usepackage{color}

C
{\lvert}{\rvert}

C
{\langle}{\rvert}

C
{.}{\rvert}

C
{\langle}{\rangle}

\DeclarePairedDelimiter{\kant}              
{[}{]}

\DeclarePairedDelimiter{\ket}               
{\lvert}{\rangle}

C
{\lVert}{\rVert}

\DeclarePairedDelimiter{\para}{(}{)}        

\DeclareMathOperator{\M}{Ma}
\DeclareMathOperator{\rey}{Re}

\newcommand{\eps}{\epsilon}

\newcommand{\f}[1]{f_i^{(#1)}}

\begin{footnotesize}

\end{footnotesize}

\newcommand{\ve}[1]{\mathbf{#1}}            

\usepackage{amssymb}

\journal{Journal of Computational Physics}

\begin{document}
\begin{frontmatter}



\title{Detailed analysis of the lattice Boltzmann method on unstructured grids}

 \author[label1]{Marek Krzysztof Misztal} 
 \ead{misztal@nbi.dk}
 \author[label1]{Anier Hernandez-Garcia}
 \ead{ahernan@nbi.ku.dk}
 \author[label1]{Rastin Matin}
 \ead{rastin@nbi.ku.dk}
 \author[label2]{Henning Osholm S\o rensen}
 \ead{osholm@nano.ku.dk}
 \author[label1]{Joachim Mathiesen}
 \ead{mathies@nbi.dk}
 
 \address[label1]{Niels Bohr Institute, University of Copenhagen, DK-2100 Copenhagen, Denmark}
 \address[label2]{Nano-Science Center, Department of Chemistry, University of Copenhagen, DK-2100 Copenhagen, Denmark}

\begin{abstract}

The lattice Boltzmann method has become a standard for efficiently solving problems in fluid dynamics. While unstructured grids allow for a more efficient geometrical representation of complex boundaries, the lattice Boltzmann methods is often implemented using regular grids. Here we analyze two implementations of the lattice Boltzmann method on unstructured grids, the standard forward Euler method and the operator splitting method. We derive the evolution of the macroscopic variables by means of the Chapman-Enskog expansion, and we prove that it yields the Navier-Stokes equation and is first order accurate in terms of the temporal discretization and second order in terms of the spatial discretization. Relations between the kinetic viscosity and the integration time step are derived for both the Euler method and the operator splitting method. Finally we suggest an improved version of the bounce-back boundary condition. We test our implementations in both standard benchmark geometries and in the pore network of a real sample of a porous rock.
\end{abstract}


\end{frontmatter}

\section{Introduction}

Based on the Boltzmann equation, lattice Boltzmann (LB) schemes have become a powerful tool for simulating complex flows in two- and three-dimensional systems. In the standard LB schemes based on uniform, regular grids, the discretization of the computational domain and the discretization of particles' velocities are coupled since the spatial grid is aligned with the characteristic directions of the velocity set. Such coupled discretization poses a severe limitation when aiming at simulating flows in complex geometries, which are encountered in several engineering problems (porous flows, aerodynamics, acoustics). This is primarily due to the fact, that in order to obtain an accurate boundary representation, a high resolution grid is required, increasing the overall size of the system (the boundary representation's accuracy is on the order of $O(h)$, while the volumetric grid's size scales like $h^{-2}$ in 2D and $h^{-3}$ in 3D, where $h$ is the grid spacing). In the recent years, various types of off-lattice Boltzmann methods have been developed in order to allow for enhanced geometric flexibility of such schemes, which might challenge the standard LB methods (see for instance \cite{ubertini_bella_succi2003}, \cite{rossi_ubertini_bella_succi2005}, \cite{Ubertinimemory}, \cite{Karlinoff} and references therein). 

Our focus here is on the finite volume schemes developed in \cite{ubertini_bella_succi2003} and \cite{rossi_ubertini_bella_succi2005}. A prominent feature of these schemes is the independence of the velocity and space discretizations. \cite{ubertini_bella_succi2003} show, using the numerical dispersion relation, that their scheme does not exhibit any dispersion effects up to the third order in wave-vector space. Also, by analysing the dispersion relation, they find that the kinematic viscosity is given by $\nu=c^2_s\tau$, indicating that numerical viscosity effects are absent (with the exception of the numerical diffusion proportional to the square of the grid spacing). This fact, as pointed out in \cite{rossi_ubertini_bella_succi2005}, requires a more careful theoretical examination. We have addressed this standing problem by means of the Chapman-Enskog expansion. Our results, as demonstrated later, corroborate those findings for the forward Euler time integration. As stated in \cite{rossi_ubertini_bella_succi2005} lack of numerical viscosity implies no mesh limitations on the highest Reynolds number that can be simulated, nonetheless, small viscosities can only be achieved with vanishingly small relaxation times. These, together with the Courant-Friedrichs-Lewy (CFL) stability condition, $\delta t<2\tau$, would imply prohibitively small time step size. However, we emphasize that this result is valid only for the forward Euler time integration and might not hold for different time integration schemes. Our analysis of the operator splitting based time integration, introduced in \cite{rossi_ubertini_bella_succi2005}, show that the kinematic viscosity is proportional to the difference between the relaxation time and the time step $\nu = c_s^2 (\tau - \delta t)$, resembling the results for the finite difference LB methods on regular grids. This interesting result might have far reaching consequences since it overcomes the constraint on the relaxation and the time step to obtain very low viscosities. 

The paper is divided in five sections and an extended appendix with details on the derivations in the main sections. In Section \ref{sec:lbm}, we introduce for completeness the basic equations for the lattice Boltzmann method. In Section \ref{sec:numerics}, we provide an overview of the implementation of the LB equation on unstructured grids and perform an analysis of the forward Euler and operator splitting temporal discretization schemes. We furthermore consider an improved version of the bounce-back boundary condition. In Section \ref{sec:exp}, we test our implementation of the LB method on a couple of benchmark systems and in the pore structure of a porous rock. The pore space of rocks is an example where the unstructured grids can provide a very efficient geometrical representation relative to the regular grids.
Porous structures are in general characterized by complex channel geometries, posing a significant challenge for most of fluid simulation software, while at the same time there is a significant industrial interest in efficient simulations of porous flow due to the relevance in groundwater flow, pollutant transport and oil recovery. In Section \ref{sec:conc}, we make a few concluding remarks. In the appendix, details can be found on the properties of the numerical scheme that we introduce as well as detail on the Chapman-Enskog expansion.

\section{Lattice Boltzmann methods}\label{sec:lbm}
The majority of lattice Boltzmann methods aim at solving the lattice Boltzmann equation 
\begin{equation}
\frac{\partial f_i}{\partial t} + \mathbf{c}_i \cdot \nabla f_i = \bar{\Omega}_i, \quad \mathrm{for }\:\:i = 0, 1, \ldots, N, \label{eq:lbe}
\end{equation}
which is a discrete formulation of the Boltzmann equation, discretized in velocity domain. Here $\mathbf{c}_i$, $i = 0,1,\ldots,N$ is the discrete set of admissible particle velocities and $f_i(\mathbf{x},t) \equiv f(\mathbf{x}, \mathbf{c}_i, t)$ is the probability density function for finding a particle in a state $(\mathbf{x}, \mathbf{c}_i, t)$; this function can be used to recover the macroscopic variables of the flow, such as mass ($\rho$) or momentum ($\rho \mathbf{u}$) density  
\begin{eqnarray}
\rho &=& \sum_{i=0}^{N} f_i, \\
\rho\mathbf{u} &=& \sum_{i=0}^N \mathbf{c}_i f_i.
\end{eqnarray}  
The term $\mathbf{c}_i \cdot \nabla f_i$ is responsible for advection of particles, and is often referred to as the \emph{streaming} term. The right-hand side of Eq. \eqref{eq:lbe}, $\bar{\Omega}_i$, is the discrete \emph{collision} operator. A popular choice is the single-relaxation Bhatnagar-Gross-Krook (BGK) operator, \cite{original_BGK}
\begin{equation}
\bar{\Omega}^{\mathrm{BGK}}_i = -\frac{1}{\tau}\left( f_i - f_i^{eq}\right), \label{eq:collision_op}
\end{equation}
where $\tau$ is the relaxation time (related to the fluid's kinematic viscosity), and $f_i^{eq}$ is the local equilibrium distribution, typically in a form of the second order expansion (third order accurate with respect to the Mach number)
\begin{equation}
f_i^{eq} = w_i \rho \left( 1+ \frac{\mathbf{c}_i \cdot \mathbf{u}}{c_s^2} + \frac{\left(\mathbf{c}_i \cdot \mathbf{u}\right)^2}{2c_s^4} - \frac{u^2}{2c_s^2}\right), \label{eq:equilibrium}
\end{equation}
where $w_i$, $i = 0,1,\ldots,N$ are the weights associated with the velocities $\mathbf{c}_i$, and $c_s$ is the lattice speed of sound. Like most authors, we use $c_s = 1/\sqrt{3}$. It has be shown by \cite{Benzi}, by means of the Chapman-Enskog expansion, that the macroscopic variables derived from Eq. \eqref{eq:lbe} yield the weakly compressible Navier-Stokes, as long as the discrete velocity layout $\mathbf{c}_{i=0,1,\ldots,N}$ and the collision operator \eqref{eq:collision_op}-\eqref{eq:equilibrium} fulfill the mass and moment conservation rules
\begin{eqnarray}
\sum_i \bar{\Omega}_i &=& 0, \\
\sum_i \mathbf{c}_i \bar{\Omega}_i &=& \mathbf{0}.
\end{eqnarray}

In this paper we present an unstructured (tetrahedral) grid based, finite volume implementation of the lattice Boltzmann method. Our work builds on previous works by \cite{ubertini_bella_succi2003, ubertini_succi_bella2004, rossi_ubertini_bella_succi2005}. We expand on their approach by introducing new solid, inlet and outlet boundary condition, which enable efficient simulations of flows in complex geometric domains, such as porous structures. Furthermore we study two different time discretization schemes, and perform the full multiscale analysis of the numerical scheme, which yield kinematic viscosities of $c_s^2 \tau$ for the forward Euler method and $c_s^2 (\tau-\delta t)$ for the operator splitting method.

\section{Numerical method} \label{sec:numerics}
\subsection{Spatial discretization of the lattice Boltzmann equation}
\label{sec:spatial_disc}
\begin{figure}
\centering
\includegraphics[width=0.66\textwidth]{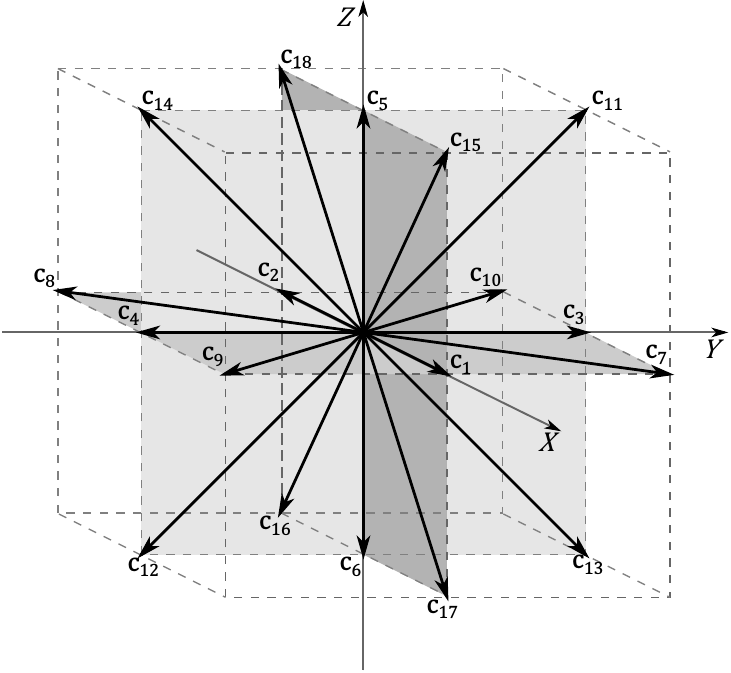}
\caption{The discrete set of admissible particle velocities in the D3Q19 layout. This set also includes $\mathbf{c}_0 = \left(0,0,0\right)$.} \label{fig:d3q19}
\end{figure}

The lattice Boltzmann equation with the collision term modeled by BGK approximation is typically discretized in the velocity domain as
\begin{equation}
\frac{\partial f_i\left(\mathbf{x}, t\right)}{\partial t} + \mathbf{c}_i \cdot \nabla f_i\left(\mathbf{x}, t\right) = -\frac{1}{\tau}\left( f_i\left(\mathbf{x},t\right) - f_i^{eq}\left(\mathbf{x}, t\right)\right), \quad i = 0, 1, \ldots, N,\label{eq:lbeq_general}
\end{equation}
where $f_i(\mathbf{x}, t)$ is the probability distribution function, $f_i^{eq}(\mathbf{x}, t)$ is the equilibrium probability distribution, $\tau$ is the relaxation time, and $\{ \mathbf{c}_i \}_{ i = 0, 1, \ldots, N }$ is the discrete set of admissible particle velocities. For the latter, we use the popular D3Q19 layout (see Fig. \ref{fig:d3q19}).

\begin{figure}
\centering
\includegraphics[width=0.4\textwidth]{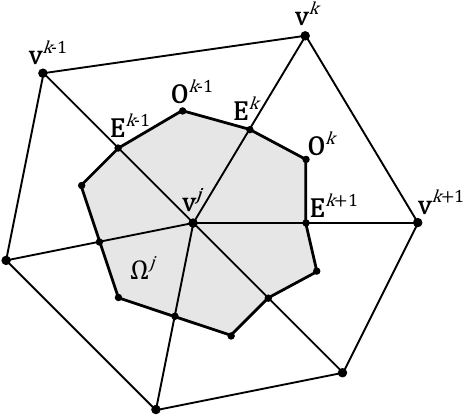}\hspace{20pt}
\includegraphics[width=0.49\textwidth]{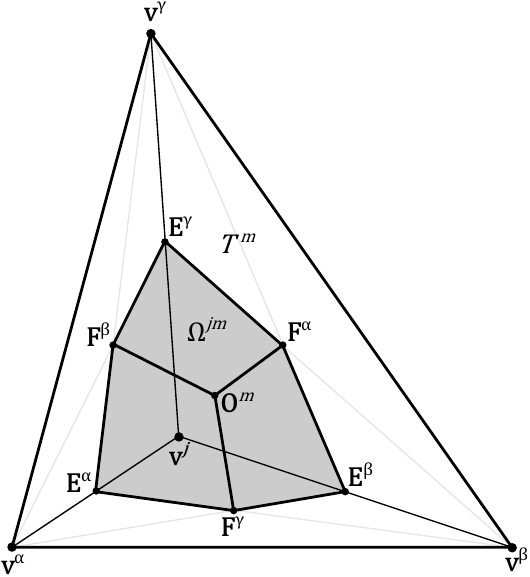}
\caption{To the left: the control volume $\Omega^j$ centered at the vertex $\mathbf{v}^j$ in a two-dimensional, unstructured grid. The control volume is constructed by connecting the barycenters $\mathbf{O}^k$ of each triangle adjacent to $\mathbf{v}^j$ with the barycenters (midpoints) $\mathbf{E}^k$ of the edges adjacent to $\mathbf{v}^j$. To the right: a contribution $\Omega^{jm}$ to the control volume $\Omega^j$ from tetrahedron $T^m$. $\Omega^{jm}$ is the convex hull of $\mathbf{v}^j$, the barycenter $\mathbf{O}^m$ of $T^m$ and the barycenters of all the edges ($\mathbf{E}^{\alpha,\beta,\gamma}$) and faces ($\mathbf{F}^{\alpha,\beta,\gamma}$) of $T^m$.} \label{fig:control_volumes}
\end{figure}

The computational domain is approximated with an unstructured, tetrahedral mesh, conforming to the solid boundary. We discretize Eq. \eqref{eq:lbeq_general} using a linear, vertex-centered, unstructured finite volume method, following \cite{ubertini_bella_succi2003,ubertini_succi_bella2004,rossi_ubertini_bella_succi2005}. This means that the probability distribution functions are defined at the vertices of the mesh 
\begin{equation}
f_i(\mathbf{v}^j,t) = f_i^j(t)
\end{equation}
and are linearly interpolated elsewhere
\begin{equation}
f_i(\mathbf{x}, t) = \sum_j f_i^j(t)\,\phi^j(\mathbf{x}), 
\end{equation}
where $\phi^j(\mathbf{x})$ is the linear interpolant function associated with vertex $\mathbf{v}^j$ (or the barycentric coordinate function, when restricted to a single element); $\phi^j(\mathbf{v}^j) = 1$, $\phi^j(\mathbf{v}^k) = 0$ for $k \neq j$, and $\phi^j$ is linear over each tetrahedron.

For each mesh vertex $\mathbf{v}^j$ we define the control volumes $\Omega^j$ as the polyhedra spanned by the barycenters of the tetrahedra, faces and edges neighbouring $\mathbf{v}^j$ (see Fig. \ref{fig:control_volumes}). By integrating Eq. \eqref{eq:lbeq_general} over $\Omega^j$ we obtain
\begin{equation}
\int_{\Omega^j}\frac{\partial f_i(\mathbf{x}, t)}{\partial t}\,d\Omega = -\int_{\Omega^j}\mathbf{c}_i \cdot \nabla f_i\left(\mathbf{x}, t\right)d\Omega - \frac{1}{\tau}\int_{\Omega^j}\left( f_i\left(\mathbf{x},t\right) - f_i^{eq}\left(\mathbf{x}, t\right) \right)d\Omega.\label{eq:weak_form_a}
\end{equation}
We approximate the left-hand side 
\begin{equation}
\int_{\Omega^j}\frac{\partial f_i(\mathbf{x}, t)}{\partial t}\,d\Omega \approx \frac{\partial f_i(\mathbf{v}^j, t)}{\partial t} V^j, 
\end{equation}
where $V^j$ is the volume of $\Omega^j$. Since $\mathbf{c}_i \cdot \nabla f_i(\mathbf{x},t) = \nabla \cdot ( \mathbf{c}_i f_i(\mathbf{x}, t) )$ we can apply the divergence theorem to the first (streaming) term on the right-hand side of \eqref{eq:weak_form_a} which yields
\begin{equation}
\int_{\Omega^j} \mathbf{c}_i \cdot \nabla f_i\left(\mathbf{x}, t\right)d\Omega = \oint_{\partial \Omega^j} \left( \mathbf{ c }_i \cdot \mathbf{ n } \right) f_i \left(\mathbf{x}, t\right) dS.
\end{equation}
Now Eq. \eqref{eq:weak_form_a} reads
\begin{equation}
\frac{\partial f_i(\mathbf{v}^j, t)}{\partial t} \approx -\frac{1}{V^j}\left( \oint_{\partial \Omega^j} \left( \mathbf{ c }_i \cdot \mathbf{ n } \right) f_i \left(\mathbf{x}, t\right) dS + \frac{1}{\tau}\int_{\Omega^j}\left[ f_i\left(\mathbf{x},t\right) - f_i^{eq}\left(\mathbf{x}, t\right) \right]d\Omega \right).
\end{equation} 
We can split the streaming term into a sum of integrals over sub-surfaces $\partial \Omega^{jm} = \partial \Omega^{j} \cap T^m$ contained in each tetrahedron $T^m$ adjacent to $v^j$
\begin{equation}
\frac{1}{V^j} \oint_{\partial \Omega^j} \left( \mathbf{ c }_i \cdot \mathbf{ n } \right) f_i \left(\mathbf{x}, t\right) dS = \sum_m \frac{1}{V^j} \oint_{\partial \Omega^{jm}} \left( \mathbf{ c }_i \cdot \mathbf{ n } \right) f_i \left(\mathbf{x}, t\right) dS. \label{eq:int_streaming}
\end{equation}
As shown in Fig. \ref{fig:control_volumes}, $\partial \Omega^{jm}$ is the union of three quadrilaterals $Q^{\gamma} = O^m F^{\alpha} E^{\gamma} F^{\beta}$, $Q^{\alpha} = O^m F^{\beta} E^{\alpha} F^{\gamma}$ and $Q^{\beta} = O^m F^{\gamma} E^{\beta} F^{\alpha}$ and it is easy to show that each of these quadrilaterals is planar. Hence, the integrals on the left hand side of Eq. \eqref{eq:int_streaming} can be simplified further as
\begin{equation}
\frac{1}{V^j} \oint_{\partial \Omega^{jm}} \left( \mathbf{ c }_i \cdot \mathbf{ n } \right) f_i \left(\mathbf{x}, t\right) dS = \frac{1}{V^j}\sum_{l = \alpha, \beta, \gamma} \left( \mathbf{ c }_i \cdot \mathbf{ n }^l \right) \oint_{Q^l} f_i \left(\mathbf{x}, t\right) dS. \label{eq:int_streaming}
\end{equation}
Recall that $f_i$ is linear within $T^m$; then, the remaining integral can be evaluated analytically and written as a linear combination of the values of $f_i$ at $\mathbf{v}^{j}$, $\mathbf{v}^{\alpha}$, $\mathbf{v}^{\beta}$ and $\mathbf{v}^{\gamma}$. That means, we can write the whole streaming term as a linear combination of values of $f_i$ at $\mathbf{v}^j$ and its direct neighbors:
\begin{equation}
\frac{1}{V^j} \oint_{\partial \Omega^j} \left( \mathbf{ c }_i \cdot \mathbf{ n } \right) f_i \left(\mathbf{x}, t\right) dS = \sum_{\mathbf{v}^k \in \mathcal{N}^j} S_i^{jk} f_i\left(\mathbf{v}^k, t\right),
\end{equation}
where $\mathcal{N}^j$ is the set containing $\mathbf{v}^j$ and all mesh vertices connected to $\mathbf{v}^j$ by a single edge, and the coefficients $S_i^{jk}$ depend only on the local mesh geometry. Notice that $S_i^{jk} \neq 0$ only if vertices $j$ and $k$ share an edge (are in each other's direct neighborhood). Then, by substituting a constant function $f_i(\mathbf{x}, t) = 1$ we obtain the following sum rule
\begin{equation}
\sum_k S_i^{jk} = \frac{1}{V^j} \oint_{\partial \Omega^j} \left( \mathbf{ c }_i \cdot \mathbf{ n } \right) dS = 0. \label{eq:sum_rule_streaming}
\end{equation}

Similarly, we split the collision term into a sum over all $T^m$ adjacent to $\mathbf{v}^j$
\begin{equation}
\frac{1}{V^j} \int_{\Omega^j}\frac{1}{\tau}\left( f_i\left(\mathbf{x},t\right) - f_i^{eq}\left(\mathbf{x}, t\right) \right)d\Omega = \frac{1}{\tau} \sum_m \frac{1}{V^j} \int_{\Omega^{jm}}\left( f_i\left(\mathbf{x},t\right) - f_i^{eq}\left(\mathbf{x}, t\right) \right)d\Omega,
\end{equation}
where $\Omega^{jm} = \Omega^{j} \cap T^m$ (as shown in \mbox{Fig. \ref{fig:control_volumes}}). We can replace the last integral with the product of the volume $V^{jm}$ of $\Omega^{jm}$ and the value of $g_i \equiv f_i-f_i^{eq}$ evaluated at the center of mass of $\Omega^{jm}$, which can be written as a linear combination of the values of $g_i$ at the vertices of $T^m$. Note that we additionally assume here that $f_i^{eq}$ is also linear over $T^m$. Finally, we can write the collision term as
\begin{equation}
\frac{1}{V^j} \int_{\Omega^j}\frac{g_i\left(\mathbf{x},t\right)}{\tau} d\Omega = \frac{1}{\tau}\sum_{\mathbf{v}^k \in \mathcal{N}^j} C^{jk} g_i\left(\mathbf{v}^k, t\right), \label{eq:collision_sum}
\end{equation}
where the coefficients $C^{jk}$ can be evaluated analytically and depend only on the local mesh geometry, and do not depend on $i$; i.e. the relation \eqref{eq:collision_sum} holds for any piecewise linear function $g_i$. In particular, it holds for a constant function $g_i(\mathbf{x}, t) = \tau$, which gives us the following sum rule
\begin{equation}
\sum_k C^{jk} = \frac{1}{V^j} \int_{\Omega^j} d\Omega = 1. \label{eq:sum_rule_collision}
\end{equation} 

In the end, we obtain the spatial discretization of the form
\begin{equation}
\frac{\partial f_i\left(\mathbf{v}^j, t\right)}{\partial t} = -\sum_{\mathbf{v}^k \in \mathcal{N}^j} S_i^{jk} f_i\left(\mathbf{v}^k, t\right)-\frac{1}{\tau}\sum_{\mathbf{v}^k \in \mathcal{N}^j} C^{jk} \left( f_i\left(\mathbf{v}^k,t\right) - f_i^{eq}\left(\mathbf{v}^k, t\right) \right). \label{eq:discrete_lb_a}
\end{equation}

\subsection{Temporal discretization}
The only term left to discretized in Eq. \eqref{eq:discrete_lb_a} is the time derivative $\partial_t f_i\left(\mathbf{v}^j, t\right)$. In this paper we examine two first-order, explicit time integration schemes: the forward Euler method and the operator splitting method. For the sake of brevity we use the notation $f_i^{(n)}(\mathbf{v}^j) \equiv f_i(\mathbf{v}^j, t^n)$, where $t^n = t^0 + n \,\delta t$, $t^0$ is the time at the beginning of the simulation and $\delta t$ is the constant time step size.

The forward Euler method is commonly used for time integration of lattice Boltzmann equation, both in regular and unstructured grid based implementations. It is stable as long as the time step size fulfils the CFL condition $\delta t < 2\tau$. It yields the following numerical scheme
\begin{equation}
f_i^{(n+1)}\left(\mathbf{v}^j\right) = f_i^{(n)}\left(\mathbf{v}^j \right) - \delta t \sum_{k} S_i^{jk} f_i^{(n)}\left(\mathbf{v}^k \right) - \frac{\delta t}{\tau}\sum_{k} C^{jk} \left( f_i^{(n)}\left(\mathbf{v}^k\right) - {f_i^{eq}}^{(n)} \left(\mathbf{v}^k\right) \right). \label{eq:time_int_euler}
\end{equation} 

Another time integration scheme investigated in this paper is the explicit operator splitting method suggested by \cite{rossi_ubertini_bella_succi2005}. In this approach, the streaming and the collision terms in Eq. \eqref{eq:discrete_lb_a} are integrated separately using the forward Euler method, which yields the following numerical scheme
\begin{eqnarray}\label{eq:time_int_os}  
f_i^{\left(n+\frac{1}{2}\right)}\left(\mathbf{v}^j\right) &=& f_i^{(n)}\left(\mathbf{v}^j \right) - \delta t \sum_{k} S_i^{jk} f_i^{(n)}\left(\mathbf{v}^k \right),\\
f_i^{(n+1)}\left(\mathbf{v}^j\right) &=& f_i^{\left(n+\frac{1}{2}\right)}\left(\mathbf{v}^j \right) - \frac{\delta t}{\tau}\sum_{k} C^{jk} \left( f_i^{\left(n+\frac{1}{2}\right)}\left(\mathbf{v}^k\right) - {f_i^{eq}}^{\left(n+\frac{1}{2}\right)} \left(\mathbf{v}^k\right) \right), 
\end{eqnarray}
where the equilibrium distribution ${f_i^{eq}}^{\left(n+\frac{1}{2}\right)}$ is evaluated using the values $f_i^{\left(n+\frac{1}{2}\right)}$. 

One of the main findings of this paper is that in an unstructured grid based setting, the kinematic viscosity of the simulated fluid depends on the choice of the time integration method, as has been previously demonstrated for regular grid based finite volume LBMs, \cite{siboni2014}. In particular we have rigorously proven (using the Chapman-Enskog expansion) and confirmed in the experiments that both schemes yield the weakly-compressible Navier-Stokes equation (up to the second order terms); scheme  \eqref{eq:time_int_euler} with kinematic viscosity 
\begin{equation}
\nu^{\mathrm{FE}} = c_s^2 \tau, \label{eq:visc_FE}
\end{equation}
and scheme \eqref{eq:time_int_os} with kinematic viscosity 
\begin{equation}
\nu^\mathrm{OS} = c_s^2 \left( \tau - \delta t \right). \label{eq:visc_OS}
\end{equation} 
The latter value stands in contrast to the value $\nu^\mathrm{OS} = c_s^2 \tau$ reported by \cite{rossi_ubertini_bella_succi2005}\footnote{In this work the time step $\delta t=\tau/20$ was sufficiently low to safely neglect the $\delta t$-shift.}. Note that this puts an additional constraint on the time step, which now reads $\delta t < \tau$.

Since, to the authors' knowledge, there is no prior, published work on the numerical analysis of unstructured grid based lattice Boltzmann methods using Chapman-Enskog expansion, we present the full proof in \ref{app:appendixA}.
\subsection{Solid boundary conditions}
\label{sec:solid_bc}
\begin{figure}
\centering
\includegraphics[width=0.4\textwidth]{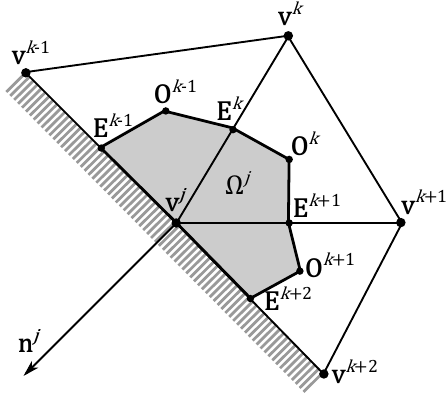}\hspace{20pt}
\includegraphics[width=0.49\textwidth]{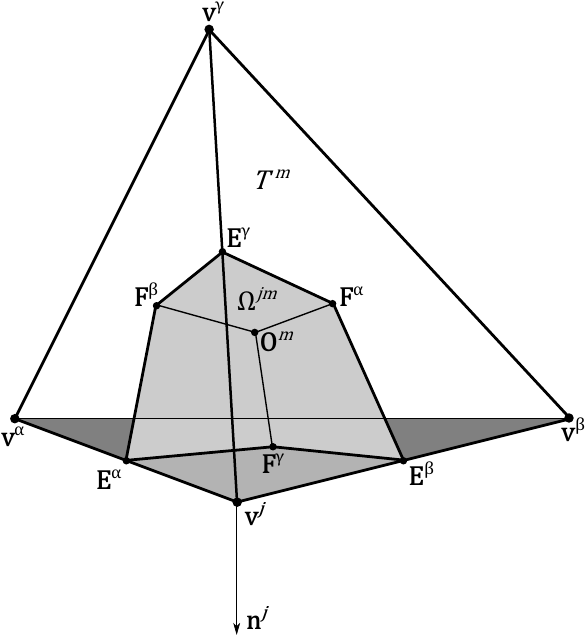}
\caption{To the left: the control volume $\Omega^j$ corresponding to the vertex $\mathbf{v}^j$ on the solid boundary. In order to evaluate streaming over a closed surface, the boundary $\partial \Omega^j$ is augmented by two solid boundary segments $\mathbf{E}^{k-1} \mathbf{v}^j$ and $\mathbf{v}^j \mathbf{E}^{k+2}$. To the right: a contribution $\Omega^{jm}$ to the control volume $\Omega^j$ from tetrahedron $T^m$, whose face $\mathbf{v}^\alpha \mathbf{v}^\beta \mathbf{v}^j$ lies on the solid boundary. The control volume boundary contribution $\partial \Omega^{jm} = \partial \Omega^j \cap T^m$ now also contains the quadrilateral $\mathbf{v}^j \mathbf{E}^\alpha \mathbf{F}^\gamma \mathbf{E}^\beta$. In both cases $\mathbf{n}^j$ refers to the normal vector to the solid boundary at $\mathbf{v}^j$.} \label{fig:control_volumes_bnd}
\end{figure}

The spatial discretization of the lattice Boltzmann equation derived in Section \ref{sec:spatial_disc} only considered bulk vertices. In this section we will discuss the solid boundary conditions and how they are included the numerical scheme. 

Accurate treatment of complex boundary conditions is non-trivial in regular grid based approaches, and while the popular bounce-back method is preferred due to its mass conservation and simple enforcement of the no-slip conditions, it typically has to be augmented with some variation of the immersed boundary method to avoid staircase artefacts, \cite{Pan2006898}. In contrast, using unstructured meshes allows us to locate the vertices of the computational grid  precisely at the physical boundary of the domain, providing us with an accurate representation of the boundary, both in terms of its geometry and topology. The adaptiveness property of unstructured meshes enables a faithful representation of the fine details of the boundary (e.g. bumps, roughness) without blowing up the overall size of the volumetric mesh.   

Like the earlier works on unstructured grid based lattice Boltzmann methods: \cite{ubertini_bella_succi2003, ubertini_succi_bella2004, rossi_ubertini_bella_succi2005, chew_shu_peng_2002}, we incorporate the solid boundary into the finite volume integration scheme via the half-covolume method. This way, the discretization of the collision term remains essentially unchanged, however, in order to properly integrate the streaming flux through a control volume $\Omega^j$ corresponding to a boundary vertex $\mathbf{v}^j$, we have to ensure that we integrate $f_i \mathbf{c}_i$ over a closed volume. We can do that by augmenting the surface $\partial \Omega^j$ constructed in the Section \ref{sec:spatial_disc} with appropriate subsets (segments in 2D, quadrilaterals in 3D) of the boundary elements, as shown in the Fig. \ref{fig:control_volumes_bnd}. In other words, we compute the streaming flux through the full topological boundary of $\Omega^j$. Notice that this preserves the sum rules \eqref{eq:sum_rule_streaming} and \eqref{eq:sum_rule_collision}.

The half-covolume method is not sufficient to enforce the appropriate solid boundary conditions (typically, the no-slip boundary conditions), since it does not provide the correct values of $f_i$ for the directions pointing from the exterior into the bulk (fluid), as pointed out in \cite{Leveque_FV, chew_shu_peng_2002}. \cite{rossi_ubertini_bella_succi2005} augment it with setting the equilibrium distribution function $f_i^{eq}$ corresponding to velocity $\mathbf{u} = \mathbf{0}$ at the boundary nodes. However, this solution does not ensure mass conservation and in our earlier experiments it destabilized the method when applied to complex solid boundaries. \cite{chew_shu_peng_2002}, in their 2D finite volume LBM, combine the half-covolume method with the bounce-back method, which ensures both no-slip boundary conditions and mass conservation. They utilize an analytical description of the solid boundary to determine which values of $f_i$ are unknown. However, in many cases (e.g. porous geometries obtained from x-ray tomography of real samples) such description is not readily available. Here we describe a completely general way of combining the bounce-back rule with half-covolume method for arbitrary, 3D solid boundaries represented by unstructured meshes.

Firstly, in the pre-processing step, we evaluate the normal vectors at all boundary vertices. Each normal vector $\mathbf{n}^j$ is approximated with an area-weighted sum of the outside-pointing normals to all boundary faces adjacent to $\mathbf{v}^j$. If these faces are nearly co-planar, the normal $\mathbf{n}^j$ is sufficient to determine which directions are unknown. We can then apply the bounce-back rule on the non-equilibrium distributions by testing whether $\mathbf{c}_{2k-1} \cdot \mathbf{n}^j < 0$ for $k = 1,\ldots,9$. If that is the case, then $f_{2k-1}$ is the unknown value and since $\mathbf{c}_{2n} \cdot \mathbf{n}^j = -\mathbf{c}_{2k-1} \cdot \mathbf{n}^j > 0$, so we can perform the substitution
\begin{equation}
f_{2k-1}(\mathbf{v}^j, t^n) = f_{2k}(\mathbf{v}^j, t^n).
\end{equation}
Alternatively if $\mathbf{c}_{2k} \cdot \mathbf{n}^j < 0$, then $\mathbf{c}_{2k-1} \cdot \mathbf{n}^j > 0$ and we perform the substitution
\begin{equation}
f_{2k}(\mathbf{v}^j, t^n) = f_{2k-1}(\mathbf{v}^j, t^n).
\end{equation}

\begin{figure}
\centering
\includegraphics[width=0.6\textwidth]{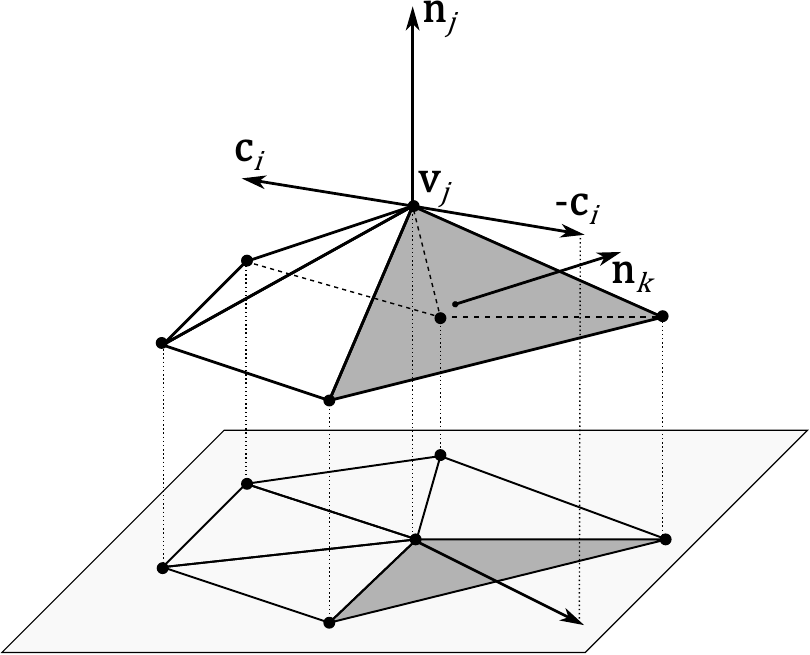}
\caption{In order to determine whether the value of $f_i$ given by the half-covolume method at a boundary vertex $\mathbf{v}_j$ is known, we test whether the vector $-\mathbf{c}_i$ is contained within the cone spanned by the boundary faces and edges adjacent to $\mathbf{v}_j$. Firstly we find out onto which boundary face $-\mathbf{c}_i$ projects; the simplest way of determining that is by projecting $-\mathbf{c}_i$, and the local neighbourhood of $\mathbf{v}_j$ onto a plane orthogonal to the normal vector $\mathbf{n}_j$. Having found such face $F_k$, we evaluate the dot product $\mathbf{c}_i \cdot \mathbf{n}_k$, where $\mathbf{n}_k$ is the outward-pointing normal vector to $F_k$. If that dot product is positive, then $-\mathbf{c}_i$ is contained within the cone, and the value $f_i$ is known; otherwise -- it is unknown. } \label{fig:bounce_back_test}
\end{figure}

However, such test is only sufficient if the solid boundary is smooth. In the general case, when the solid boundary contains sharp, non-smooth details, a more refined procedure has to be applied to determine whether a value $f_i$ is unknown. This is the case if the vector $-\mathbf{c}_i$ is not contained in the cone delimited by the boundary faces and edges adjacent to a boundary vertex $\mathbf{v}_j$ (see Fig. \ref{fig:bounce_back_test} for details). Notice that for every boundary vertex and every direction this test has to be performed only once per simulation. In our implementation it is performed in the pre-processing step, hence it does not affect the overall performance of the method, except for a slight memory overhead related to storing 18 binary flags per each boundary vertex, which indicate whether $f_i$, $i = 1,\ldots,18$ is known or unknown at that vertex. Then, the bounce-back rule can be applied to the non-equilibrium distributions via the following substitutions
\begin{eqnarray}
f_{2k-1}(\mathbf{v}^j, t^n) = f_{2k}(\mathbf{v}^j, t^n) &&\  \textrm{if } f_{2k-1} \textrm{ is unknown and } f_{2k} \textrm{ is known},\\
f_{2k}(\mathbf{v}^j, t^n) = f_{2k-1}(\mathbf{v}^j, t^n) &&\  \textrm{if } f_{2k} \textrm{ is unknown and } f_{2k-1} \textrm{ is known},
\end{eqnarray}  
for $k = 1,\ldots,9$. If both $f_{2k-1}$ and $f_{2k}$ are unknown, instead we substitute
\begin{equation}
f_{2k-1}(\mathbf{v}^j, t^n), f_{2k}(\mathbf{v}^j, t^n) = \frac{1}{2}\left( f_{2k-1}(\mathbf{v}^j, t^n) + f_{2k}(\mathbf{v}^j, t^n) \right).
\end{equation}
This rule proved sufficiently good in our experiments, but it could be further improved. Particularly appealing is the method proposed by \cite{Chikatamarla20131925}, used in turbulent flow simulations, where they approximate the populations in the unknown directions (which we can identify as described above) using the target values of density and velocity at the boundary node, $\rho_{\mathrm{target}}$ and $\mathbf{u}_{\mathrm{target}}$ respectively. While their method has been developed for regular grids, its generalization to unstructured grids is straightforward. 

\subsection{Inlet and outlet boundary conditions}
We enforce the pressure values at the inlet and outlet by applying bounce-back to the non-equilibrium parts of the unknown distributions after streaming, after \cite{Zou_He_BC_3D}. However, in 3D this approach, together with the closure relations for mass and momentum conservation, leads to excess momentum in the two dimensions that span the plane of the inlet or outlet. Following \cite{Zou_He_BC_3D}, we get rid of this excess momentum by redistributing it among the unknowns $f_i$ pointing into the fluid. The nodes at the boundaries of the inlet and outlet are treated as all other solid boundary nodes.  

Following \cite{rossi_ubertini_bella_succi2005}, we augment all our meshes with a certain number $N$ of additional, identical buffer layers of elements at the inlet and outlet. The purpose of these two buffers is to increase the stability of the method, as they ensure that the control volumes at the inlet and outlet nodes close up. After streaming, colliding and applying pressure boundary conditions, the values of $f_i$ at each inlet and outlet node are copied to the corresponding $N$  buffer nodes in order to enforce complete hydrodynamic equilibrium in these regions.

\subsection{Meshing considerations}
\label{sec:meshing}
It is a well-established fact, recognized by both computational fluid dynamics and computational mechanics communities, that the stability and accuracy of an unstructured grid based simulation strongly depends on the quality of the grid. Several quality measures for tetrahedral meshes have been proposed, and in isotropic case, they all tend penalize tetrahedra which significantly differ from the regular tetrahedron, for an in-depth comparison see \cite{Shewchuk02whatis}. 

The most relevant observation to our method is that elements with large dihedral angles (close to $\pi$) cause significant interpolation errors, manifesting themselves as gradient artefacts and thus should be avoided, \cite{Shewchuk02whatis}. In order to ensure that the computational domain does not contain such degenerate elements, we apply a local mesh improvement method, similar to that described by \cite{Klingner:2007:ATM}. 

Proper treatment of the solid boundary conditions sets another restriction on the mesh structure. Our method for handling such boundary conditions in Section \ref{sec:solid_bc} hinges on the fact that the half-covolume method produces correct values of the particle distribution function $f_i$ at a boundary vertex $\mathbf{v}^j$ in the directions pointing away from the fluid, and our method for identifying the unknown directions considers only the local boundary patch (i.e. the boundary faces containing given $\mathbf{v}^j$). However, if the other vertices, connected to $\mathbf{v}^j$ by a single, non-boundary edge, also lie on the solid boundary (which can be the case if the mesh is under-resolved in narrow channels), then the half-covolume method can produce erroneous values in the directions identified as known. Hence special care has to be taken when designing or optimizing the computational mesh in order to avoid such configurations. 

Finally, as shown in \ref{app:appendixA}, the linear terms contributing to numerical diffusion are on the form
\begin{equation}
\sum_k C^{jk} \mathbf{r}^{jk},
\end{equation}
where $\mathbf{r}^{jk} = \mathbf{v}^k-\mathbf{v}^j$. By definition of the collision matrix \eqref{eq:collision_sum} we have
\begin{equation}
\sum_k C^{jk} \zeta(\mathbf{v}^k) = \frac{1}{V^j} \int_{\Omega^j} \zeta(\mathbf{x})\, d\Omega,\label{eq:collision_def}
\end{equation}
where $\Omega^j$ is the control volume associated with vertex $\mathbf{v}^j$, and $\zeta(\mathbf{x})$ is an arbitrary, continuous function, linear over each element. In particular, Eq. \eqref{eq:collision_def} holds for $\zeta(\mathbf{x}) = \mathbf{r}^j(\mathbf{x}) \equiv \mathbf{x}-\mathbf{v}^j$. Hence
\begin{equation}
\sum_k C^{jk} \mathbf{r}^{jk} = \frac{1}{V^j} \int_{\Omega^j} (\mathbf{x}-\mathbf{v}^j) \, d\Omega.
\end{equation} 
It is evident that the term on the right-hand side becomes zero, if $\mathbf{v}^j$ lies in the geometric center of the control volume $\Omega^j$, i.e. when
\begin{equation}
\mathbf{v}^j = \frac{1}{V^j} \int_{\Omega^j} \mathbf{x} \, d\Omega.\label{eq:geometric_center}
\end{equation}
This implies that it is possible to remove the first-order numerical diffusion  originating from spatial discretization, by designing or optimizing the mesh in a way that places each vertex at the geometric center of the control volume associated with this vertex. In practice, this can be done by designing an iterative mesh smoothing procedure, i.e. a procedure for displacing mesh vertices without changing their connectivity, which aims at satisfying criterion \eqref{eq:geometric_center}.

\section{Experiments and results}\label{sec:exp}
In this section we benchmark the LBM by applying it to a freely decaying shear wave in a periodic box and to Poiseuille flow using both forward Euler and operator splitting time integration. In addition we will use these geometries to verify the derived expressions for the viscosities in both time integration schemes, \eqref{eq:visc_FE} and \eqref{eq:visc_OS} and estimate their accuracy as a function of grid resolution.

\begin{figure}
\centering
\includegraphics[width=0.49\textwidth]{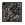}
\includegraphics[width=0.49\textwidth]{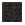}
\caption{Cross sections of the cube meshes $\mathcal{M}_1$ (left, 9273 tetrahedra) and $\mathcal{M}_2$ (right, 235447 tetrahedra) used in the freely decaying shear wave experiments, taken at $z = 0$.} \label{fig:cubes}
\end{figure}

\subsection{Freely decaying shear waves}

 We first consider a freely decaying shear wave in a periodic box, which allows us to avoid the use of boundary conditions, which shall be analysed below by considering a Poiseuille flow. 
 
We set initially a velocity profile equal to
   \begin{equation}
   v_y(x,t=0)=v^0_y \sin(k_x x),
   \end{equation}
    in which $v_y$ denotes the velocity along the $y$ axis and $k_x$ -- the wave number. For zero pressure gradient and relatively low Reynolds numbers the Navier-Stokes equation has an analytical solution given by

      \begin{equation}
   v_y(x,t)=v^0_y \sin(k_x x)\,\mathrm{e}^{-\nu k^2_x t},
   \end{equation}
 from which we have

 \begin{equation}
 -\frac{1}{k^2_x t} \ln \left(\frac{v_y(x,t)}{v^0_y \sin(k_x x)}\right)=\nu.
\end{equation}

 Then, from the measurement of the time series of the velocity at a certain point we can obtain the viscosity of the simulated fluid. Specifically, if we select a point $x_o$ such that $\sin(k_x x_o)=1$ we can obtain the viscosity from the following formula

  \begin{equation}
 -\frac{1}{k^2_x t} \ln \left(\frac{v_y(x_o, t)}{v^0_y }\right)=\nu.
\end{equation}

 The simulations presented here were performed on two meshes (see Fig. \ref{fig:cubes}) whose characteristic grid spacing (measured as the mean edge length) are $0.1$ and $0.036$ measured in LB units. We shall refer to these meshes as $\mathcal{M}_1$ and $\mathcal{M}_2$, respectively. The chosen velocity amplitude is $v^0_y=0.05$, which corresponds to $\mathrm{Ma}\approx 3\cdot 10^{-2}$. We performed a series of simulations changing the relaxation time for both forward Euler and operator splitting schemes in order to measure the viscosity and compare it with the closed form solutions given by \eqref{eq:visc_FE} and \eqref{eq:visc_OS}, respectively. 

\begin{figure}[h!]
  \centering

\includegraphics{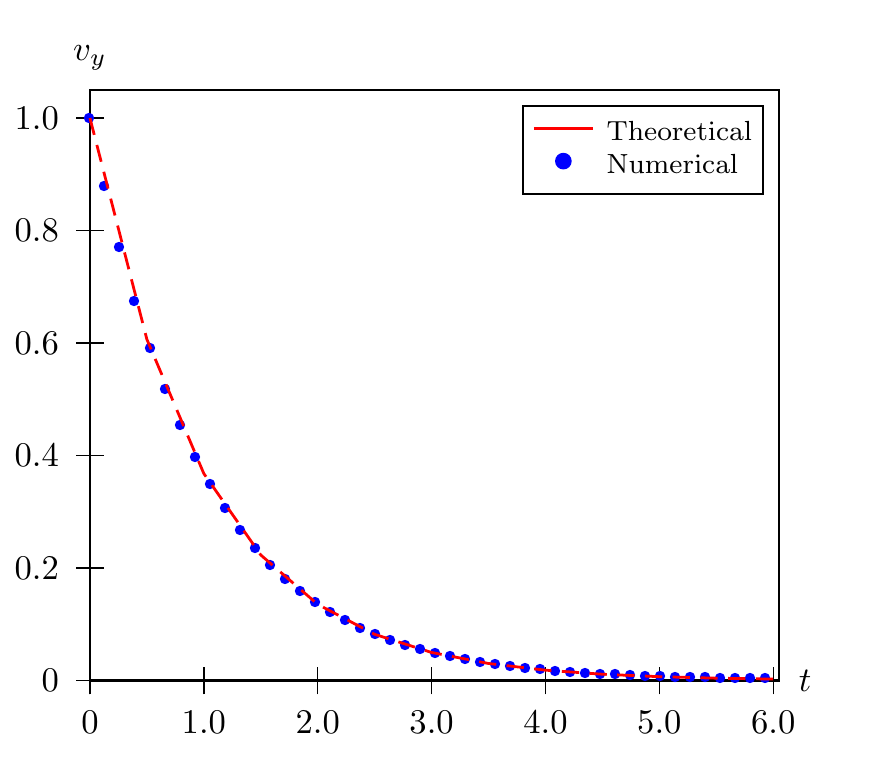}
\caption{The time evolution of normalized velocity along the $y$ axis compared against the analytic solution. In the figure the time has been rescaled by $t_o=\left(k_x^2\nu\right)^{-1}$. The spatial grid contains $235447$ elements.  The relaxation time and time step used were $0.08$ and $0.05$, respectively. From the relaxation of the velocity along the $y$ axis we obtain the kinematic viscosity $\nu = 0.02637$, which deviates by approximately $1$\% from the theoretical value according to  the relation $\nu=c^2_s\tau$. The corresponding Reynolds number $\mathrm{Re}=12$.}
\label{fig1cube32}
\end{figure}

 In Fig.~\ref{fig1cube32} we present the time series of the  velocity $v_y(x_o, t)/v_y(x_o, 0)$ versus $t\,k_x^2\,\nu$ resulting from the simulation with the forward Euler scheme in $\mathcal{M}_2$, and compare it against the analytic solution. The relaxation time and time step used were equal to $0.08$ and $0.05$, respectively, corresponding to the Reynolds number of $\text{Re}\approx 12$.  From the relaxation of the $y$-coordinate of the velocity we obtain the kinematic viscosity $\nu = 0.02637$, which deviates by approximately $1$\% from the theoretical value given by the relation \eqref{eq:visc_FE}.

  In Tables \ref{tab:visc_table_shear_flow_OS16} and  \ref{tab:visc_table_shear_flow_OS32} we present the results for the simulations with the OS time integration in both meshes for several  relaxation times and time steps. We can observe a remarkable agreement between the viscosity values determined by these numerical experiments and the theoretical values given by \eqref{eq:visc_OS}. Moreover, if we compare the fractional deviation in viscosity $\delta \nu=\lvert \frac{\nu_e-\nu_t}{\nu_t}\rvert$ in both meshes, for a given $\tau$ and $dt$, we can see that $\delta \nu$ is approximately $4$ times lower for $\mathcal{M}_2$, indicating that the error scales approximately as the square of the grid spacing $r^2$, as suggested by our Chapman-Enskog analysis.   

  \begin{table}[h!]
  \centering
  \begin{tabular}{ccccc}
	$\tau$ &$\delta t$ &$\nu_t$ &$\nu_e$ &$\delta \nu$  \\
	\hline
	\hline				
	 $0.08$    &0.04   &0.0133  &0.0132   &0.95\% \\
	 $0.08$    &0.06   &0.00667 &0.00684  &2.55\% \\
	 $0.04$    &0.02   &0.00667 &0.00661  &0.9\% \\
	 $0.04$    &0.03   &0.00333 &0.00342  &2.7\% \\
	 $0.01$    &0.005  &0.001667 &0.001668  &0.1\% \\
	\hline
  \end{tabular}
  \caption{Comparison between the numerical estimate of kinematic viscosity $\nu_e$ and the theoretical value $\nu_t = c_s^2 (\tau - \delta t)$ for several relaxation times and time step sizes. We can see from the fractional deviation in viscosity (see text for definition) a remarkable agreement between $\nu_t$ and $\nu_e$. The simulations were performed on the coarser mesh $\mathcal{M}_1$.} 
  \label{tab:visc_table_shear_flow_OS16}
\end{table}      

  \begin{table}[h!]
  \centering
  \begin{tabular}{ccccc}
	$\tau$ &$\delta t$ &$\nu_t$ &$\nu_e$ &$\delta \nu$  \\
	\hline
	\hline				
	 $0.08$    &0.04   &0.01333  &0.01328  &0.3\% \\
	 $0.04$    &0.02   &0.00667 &0.00665   &0.28\% \\
	 $0.04$    &0.03   &0.00333 &0.00336   &0.67\% \\
	 $0.01$    &0.005  &0.001667 &0.001664  &0.1\% \\
	\hline
  \end{tabular}
  \caption{Comparison between the numerical estimate of kinematic viscosity $\nu_e$ and the theoretical value $\nu_t = c_s^2 (\tau - \delta t)$ for several relaxation times and time step. We can see from the fractional deviation in viscosity a remarkable agreement between $\nu_t$ and $\nu_e$. The simulations were performed on the finer mesh $\mathcal{M}_2$.} 
  \label{tab:visc_table_shear_flow_OS32}
\end{table}      

  We have thus provided numerical evidence for the theoretical finding that the kinematic viscosity in the operator splitting scheme does depend on the time step size. 

\begin{figure}
\centering
\includegraphics[width=0.992\textwidth]{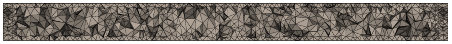}\\
\includegraphics[width=\textwidth]{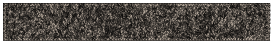}
\caption{Cross sections of the cylindrical pipe meshes used in the Poiseuille flow experiments, taken at $z = 0$, showing their internal structures. The coarse model (top) contains 32166 tetrahedra, and the fine model (bottom) contains 156749 tetrahedra. } \label{fig:pipe}
\end{figure}

\subsection{Poiseuille flow}
\begin{figure}[h!]
  \centering
    \includegraphics{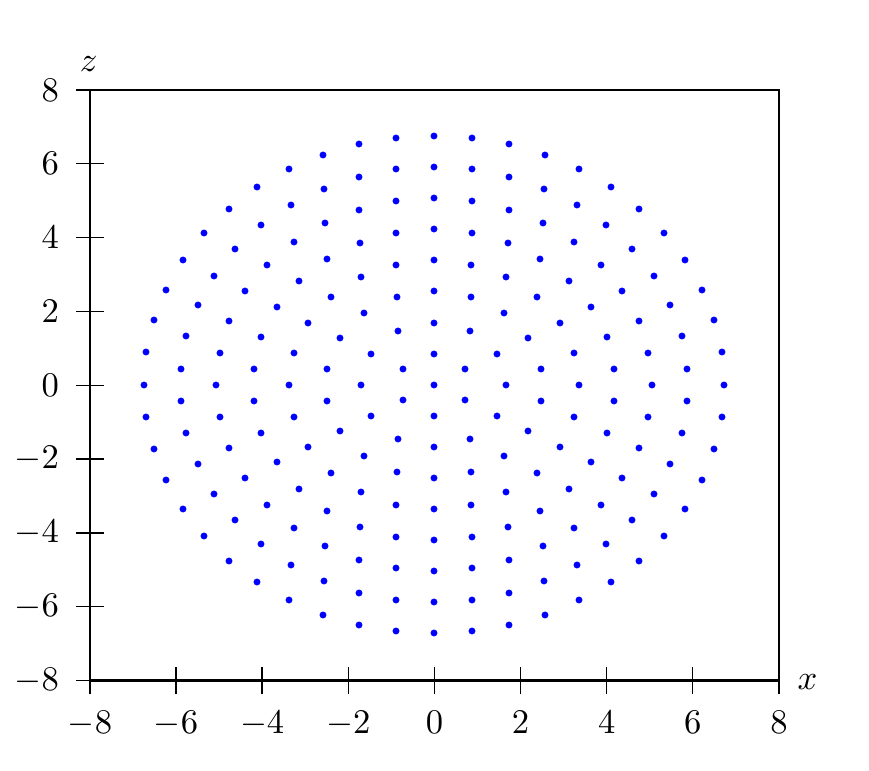}
      \caption{The inlet of the fine pipe used to model Poiseuille flow. The grid contains 217 nodes.}
      \label{fig:inlet_grid}
\end{figure}

We consider flow in a cylindrical pipe, driven by a constant volumetric force acting along the symmetry axis, with no-slip boundary conditions employed at the outer edge and periodic boundary conditions at the inlet and outlet. As illustrated in Fig. \ref{fig:pipe} and \ref{fig:inlet_grid}, the unstructured grid accurately represents the curved boundary. This geometry is particularly interesting to benchmark since it readily serves as a platform for investigating turbulent flows, e.g. by adding roughness to the boundaries.

\begin{figure}[h!]
  \centering
    \includegraphics{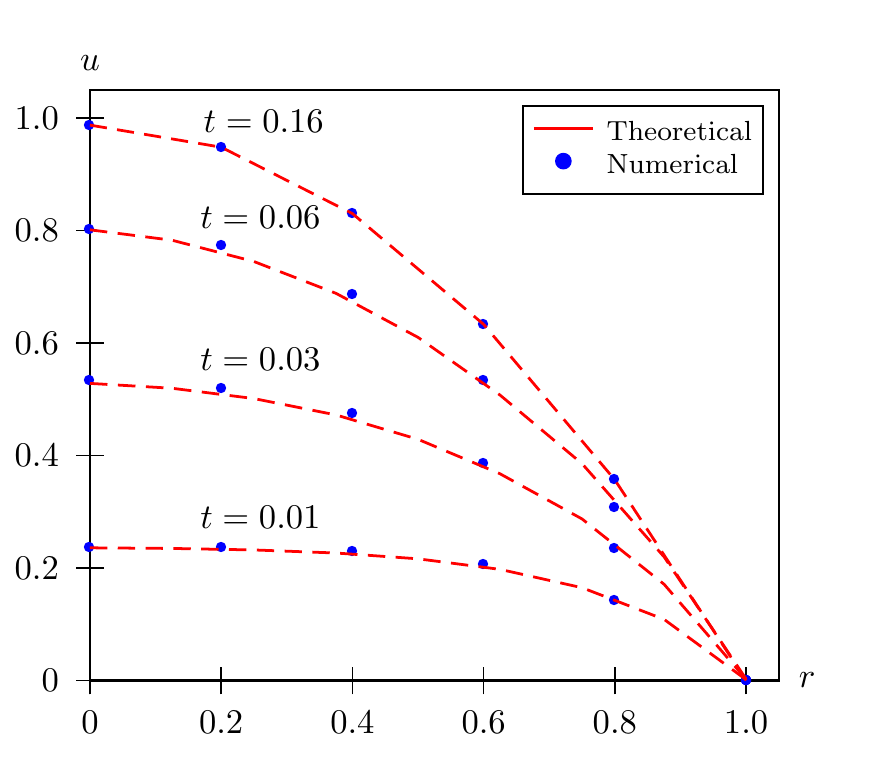}
      \caption{Convergence of the radial velocity profile towards the theoretical parabolic shape of the Poiseuille flow. The grid contains 31117 nodes and each data point is an average along the entire pipe.}
      \label{fig:increase_parabola}
\end{figure}

We drive the flow to a maximum speed $\text{Ma}=2\cdot 10^{-3}$ with $\tau = 8 \cdot 10^{-2}$ and $\delta t = 10^{-2}$, and with the pipe's radius $R=6.7$ this corresponds to $\text{Re}=0.2$. The radial velocity profile $u(r,t)$ as a function of the radius $r$ and time $t$ can be analytically obtained from the Navier-Stokes equations by utilizing the symmetry and is given by
\begin{equation}
u(r, t) = \left(1-r^2\right) - 8\sum_{n=1}^{\infty}{\lambda_n^{-3}\,\frac{J_0(\lambda_nr)}{J_1(\lambda_n)}\,\mathrm{e}^{-\lambda_n^2t/\rey}}, \quad 0\leq (u, r) \leq 1,
\end{equation}
where $J_n$ is the $n$th order Bessel function of first kind, $\lambda_n$ the $n$th positive root of $J_0$. In the above equation the velocity and radial coordinate are measured in units of the velocity at the center of the pipe and the radius, respectively. In Fig. \ref{fig:increase_parabola} we illustrate the velocity profiles obtained numerically for increasing times, showing good agreement with the corresponding analytic profiles.

We now look at the decay of a given radial velocity profile $f(r)$, which has the closed form solution
\begin{equation}
u(r, t) = \sum_{n=1}^{\infty}{\para[\bigg]{\frac{2}{J_1^2(\lambda_n)}  \int_0^1{r'f(r')J_0(\lambda_nr')dr'}} J_0(\lambda_n r)\,\mathrm{e}^{-\lambda_n^2t/\rey}}.
\end{equation}
Due to time-reversal symmetry the decay profiles are identical to those in Fig. \ref{fig:increase_parabola}. However, by taking as initial profile $f(r') = u_{\text{max}}(0) J_0(\lambda_1 r')$ we are able to find an explicit expression for the kinematic viscosity of our simulated hydrodynamics for times $t>0$
\begin{equation}
\nu = -\frac{R^2\ln(u_{\max}(t)/u_{\max}(0))}{\lambda_1^2t}. \label{eqn:visc_poiseuille}
\end{equation}
We can validate the derived expressions for the viscosity by measuring the steady-state value \eqref{eqn:visc_poiseuille} in our system. Our results are summarized in Table \ref{tab:visc_table_Poiseuille} for both the FE and OS schemes, showing the fractional error in the simulated viscosity. The results are within a few percent of the analytical solution and, as mentioned in \cite{ubertini_bella_succi2003}, the second-order effect due to numerical diffusion is found to scale inversely with the number of elements.

\begin{table}[h!]
  \centering
 \begin{tabular}{ccccccc}
    \# nodes    &$\tau$    &$\delta t$    &$\nu_t$    &$\nu_e$    &$\delta \nu$    &TS  \\
    \hline
    \hline
    7857		    &0.08     &0.04          &0.0267     &0.0238     &11.846\%         &FE \\
  		        &0.08     &0.02          &0.0267     &0.0241     &10.465\%         &FE \\
      		    &0.04     &0.02          &0.0133     &0.0121     &10.106\%         &FE \\  \hline
    31117	    &0.08     &0.04          &0.0267     &0.0262     &1.589\%          &FE \\
         	    &0.08     &0.02          &0.0267     &0.0262     &1.621\%          &FE \\
         	    &0.04     &0.02          &0.0133     &0.0131     &1.549\%          &FE \\
    \hline  
    \hline      
    7857 	    &0.08     &0.04          &0.0133     &0.0128     &4.544\%          &OS \\
         	    &0.08     &0.02          &0.0200     &0.0185     &8.356\%          &OS \\
         	    &0.04     &0.02          &0.0067     &0.0064     &4.195\%          &OS \\ \hline           
    31117 	    &0.08     &0.04          &0.0133     &0.0132     &0.658\%          &OS \\                
         	    &0.08     &0.02          &0.0200     &0.0197     &1.321\%          &OS \\                    
         	    &0.04     &0.02          &0.0067     &0.0066     &0.732\%          &OS \\
    \hline
 \end{tabular} 
  \caption{Fractional deviation in viscosity $\delta \nu$ for two different meshes and two different time-stepping (TS) schemes. The maximum velocity $u_{\max}(t)$ is obtained by averaging around the symmetry axis throughout the whole pipe.}
  \label{tab:visc_table_Poiseuille}
\end{table}

\begin{figure}
\centering
\includegraphics[width=0.49\textwidth]{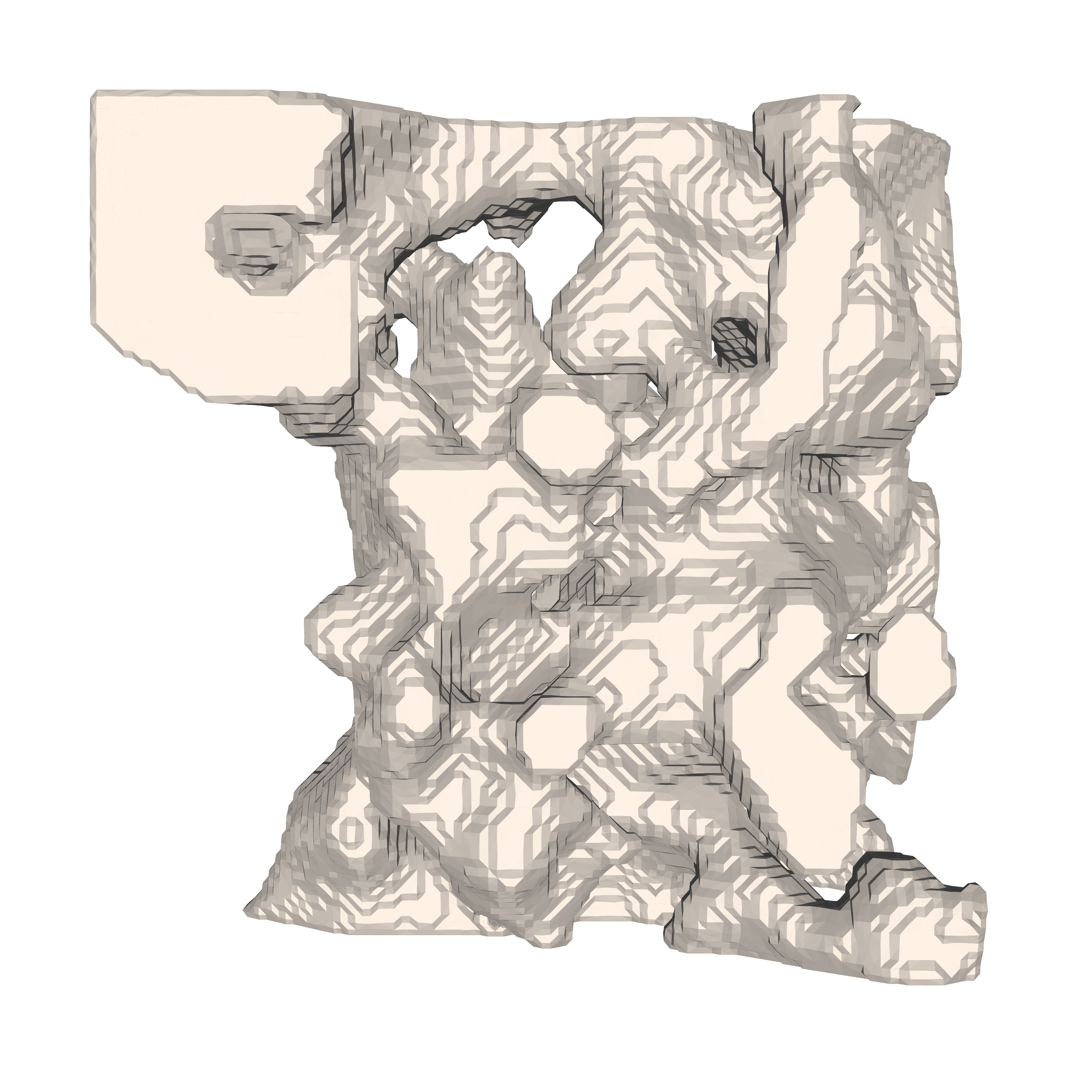}
\includegraphics[width=0.49\textwidth]{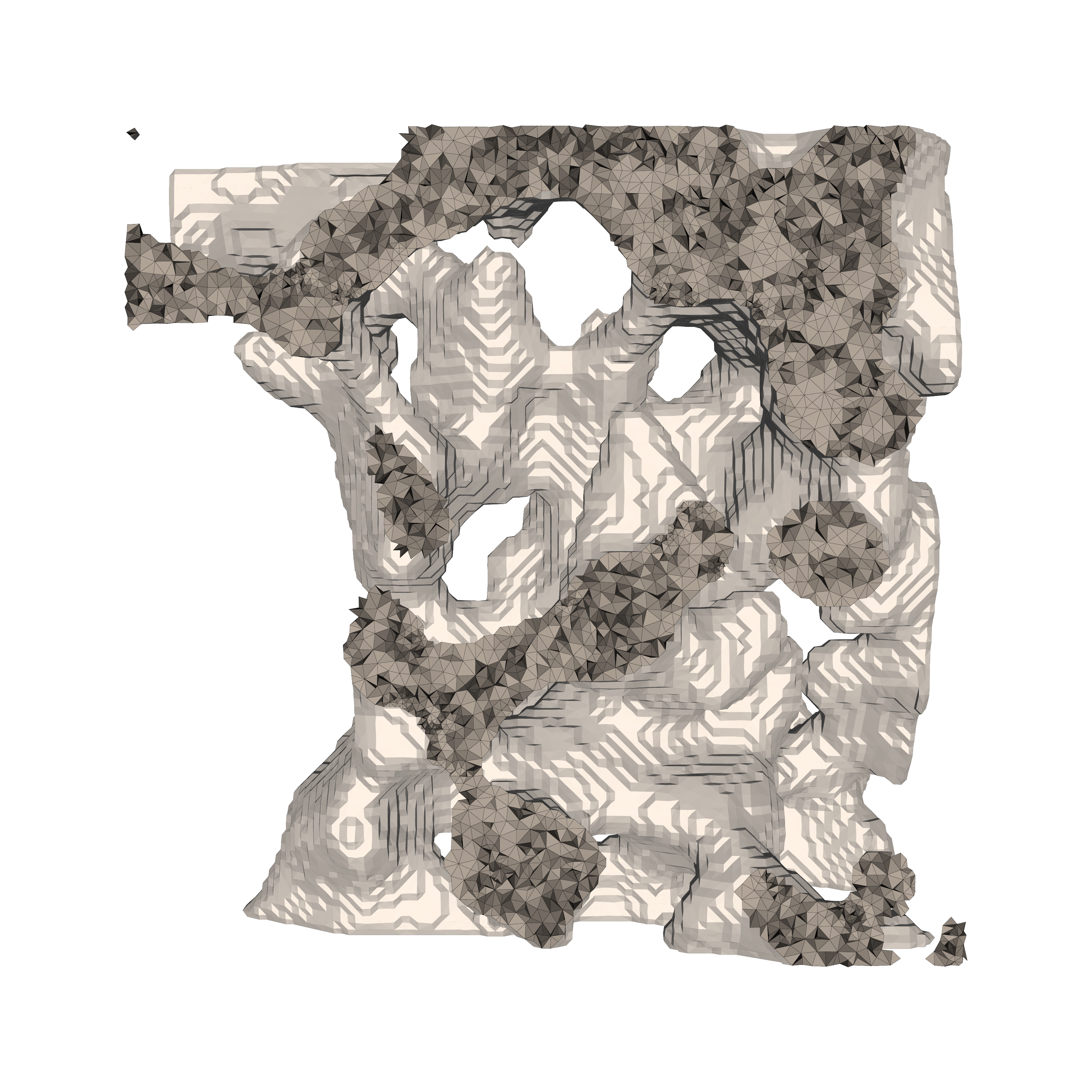}\\
\includegraphics[width=0.49\textwidth]{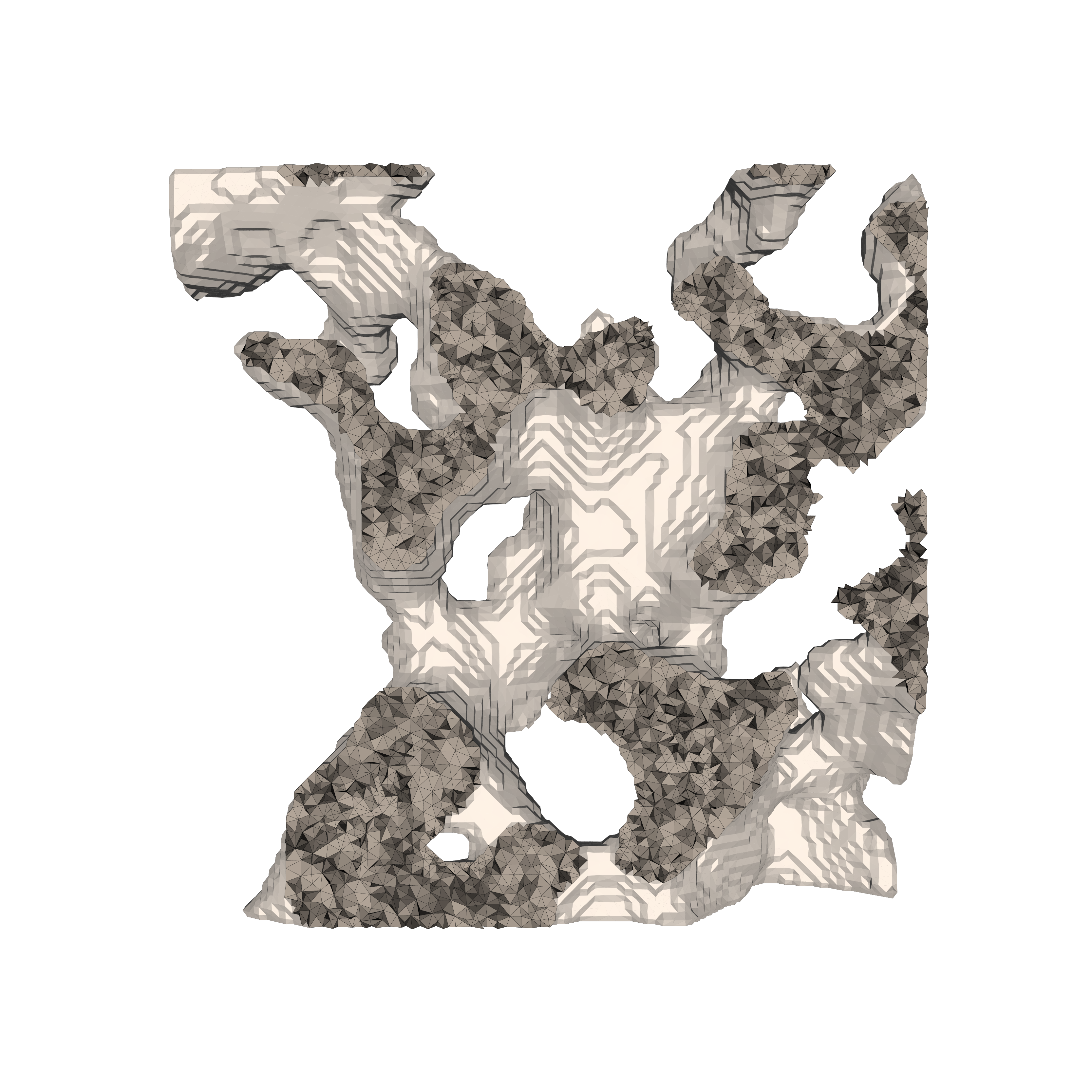}
\includegraphics[width=0.49\textwidth]{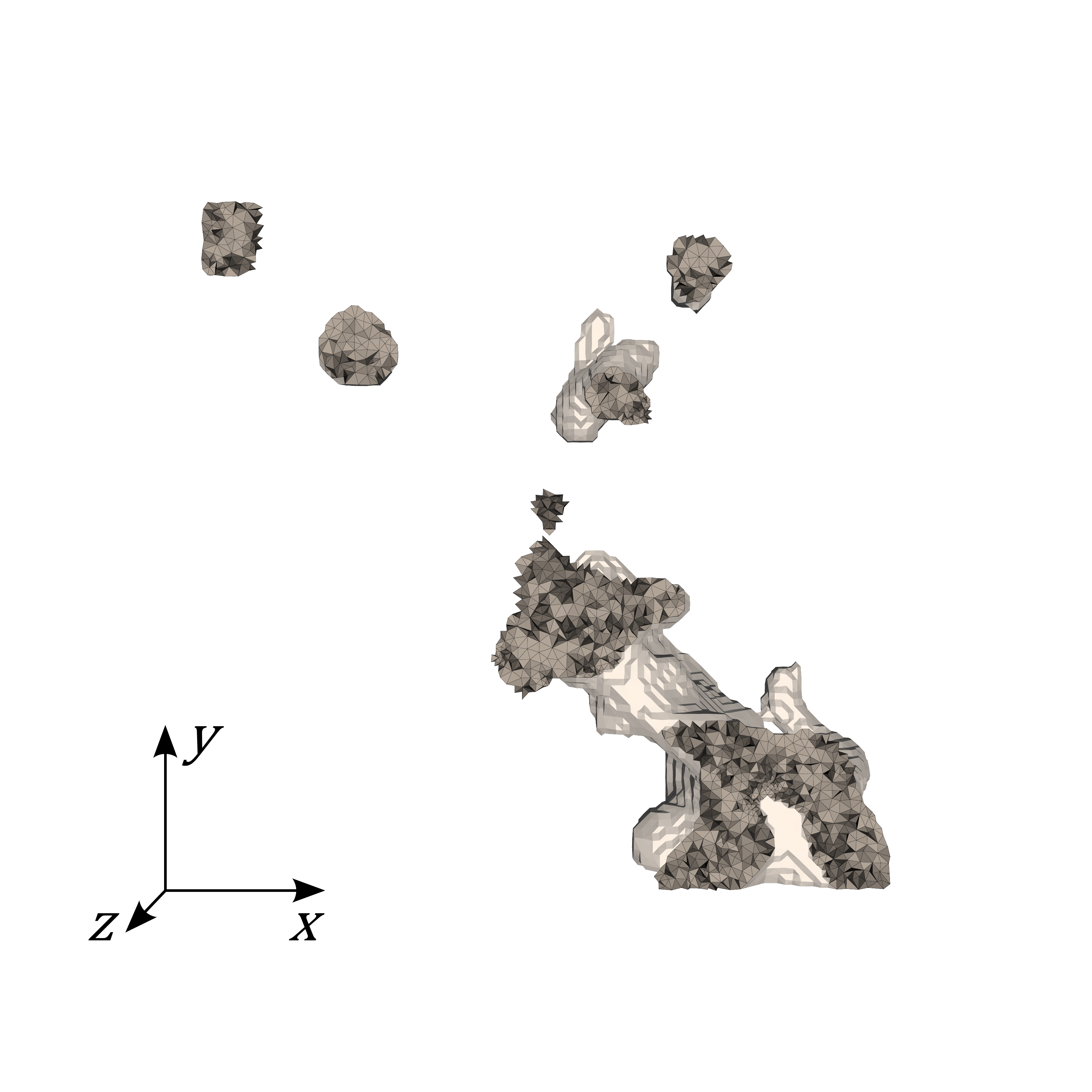}
\caption{The porous sample used in the experiment together with cross sections taken at 75\%, 50\% and 25\% of the sample's depth, showing the internal mesh structure. The tetrahedral mesh contains 621818 elements.} \label{fig:pores1}
\end{figure}

\subsection{Flow in a porous sample}
Accurate calculation of single phase flow permeability through complex pore networks in porous media is important for many industrial and scientific applications. Therefore, we have  tested our finite volume implementation of the LBM on a subvolume of  the real natural porous material of an outcrop of bryozoan chalk from R{\o}dvig, Denmark. The digital 3D image was obtained by X-ray nanotomography \cite{cloetens} measured at beamline  ID22 at the European Synchrotron Radiation Facility, France. The  reconstructed volume had a voxel size of 25 nm and an optical resolution about 150 nm. Details about the data collection and reconstruction can be found in \cite{muter2014}. The reconstructed images were corrected for ring artefacts before segmentation by a dual filtering and Otsu thresholding procedure \cite{muter2012}. For the LBM calculations we used a subvolume of 100$^3$ voxels, which gives a side length of 2.5 microns (Fig. \ref{fig:pores1}).

In addition to the benchmarks in the previous subsections, we further test our finite volume implementation of the LBM in the pore space of a limestone sample. The pore space is obtained by a computed tomography with a 25 nanometer voxel size resolution and the total linear size of the sample is approximately 3 microns. In many industrial and scientific settings, it is important to determine the single phase flow permeability. The unstructured grid, considered here, allows for a relatively simple geometrical representation of the complex pore space and therefore reduces the number of computational elements needed relative to the LBM formulated on regular grids. 

In order to drive the fluid in the sample, we impose a pressure difference between the inlet and outlet planes. The inlet and outlet conditions were implemented as described in \cite{Zou_He_BC_3D}. In our calculations, we introduce a flow in the direction perpendicular to the inlet plane along the $y$-axis.   

The permeability is determined from the empirical Darcy's law, which states that under steady-state flow conditions, the flow rate through a cross section $Q$ is proportional to the pressure drop $\Delta P$ that drives the fluid, 
\begin{equation}
Q = \int{\ve u \cdot d\ve A} = -\frac{kA}{\mu}\, \frac{\Delta P}{L},
\label{eqn:Darcy}
\end{equation}
where $k$ is the permeability, $\mu$ is the dynamic viscosity of the fluid, $\ve u$ the velocity, $A$ the cross-section of the medium and $L$ is the distance between the inlet and outlet planes. 

 In principle, the permeability is a tensorial quanitity, since different flow permeabilities might be achieved if different inlet planes are chosen. Here we have constructed the mesh such that the rock is impermeable in the directions orthogonal to the outlet plane normal (i.e.~there is no net flux in the \textit{x} and \textit{z} directions) and we therefore only determine the component $k_{yy}$. The other permeability components are easily achieved by a simple change of the inlet and outlet planes.

For a given time step $\delta t$, relaxation time $\tau$ and inlet (outlet) pressure $\rho_I$  ($\rho_O$), we determine the steady-state flow rate by averaging over the faces $i$ lying on the outlet, $Q \approx \sum_i \left\langle v_i \right\rangle A_i$. For consistency we check that $Q_O=Q_I$. 
 
In our model, we define the dimensionless permeability as 
\begin{equation}
k^*=\frac{k}{A},
\label{permdimensionless}
\end{equation}
which is, consequently, only a function of the Reynolds number and independent of the system of units we use to measure it. Substitution of \eqref{permdimensionless} in \eqref{eqn:Darcy} yields 
\begin{equation}
Q = -\frac{k^*A^2}{\mu}\frac{\Delta P}{L},
\label{eqn:Darcydimesionless}
\end{equation}
We determine the value of $k^*$ from \eqref{eqn:Darcydimesionless} expressing all the magnitudes in LB units. From this, we can obtain the value of the permeability in any arbitrary system of units $k_s$ according to the relation
\begin{equation}
k_s=k^*A_s
\label{eqn:sutransform}
\end{equation}
 For the Poiseuille flow we have $k^*=\frac{\pi}{8}$, a result that was readily benchmarked in our pipe flow simulations.  


\begin{figure}[h!]
\centering
\includegraphics[width=0.66\textwidth]{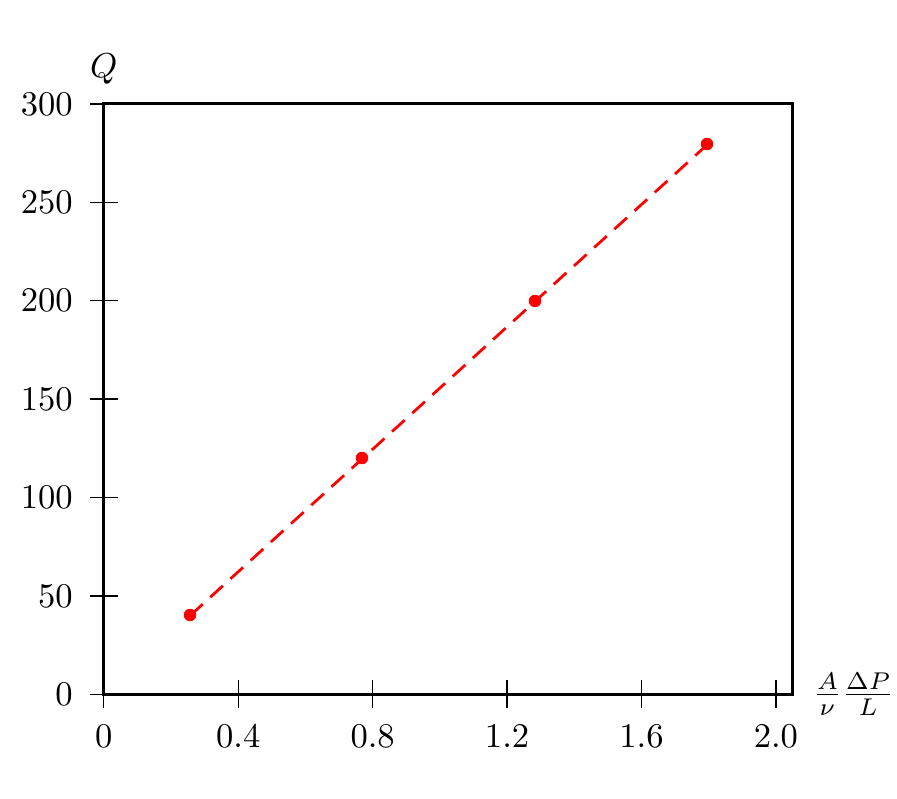}
\caption{The outlet flow rate $Q$ versus pressure difference for our porous medium, displaying a linear relationship in accordance with Darcy's law.  The presented data is in LB-units. From these measurements we estimated the permeability as $k_P = 6.5$ mD.}
\label{fig:find_perm}
\end{figure}

\subsubsection{Multiple relaxation time LBM}
The single-relaxation BGK model introduced in Section \ref{sec:numerics} suffers from viscosity-dependent flow as the fluid-solid location depends on the relaxation $\tau$ when bounce-back is employed, \cite{Pan2006898}. To circumvent this issue we employed the multi-relaxation model, \cite{el00312}. We start by generalizing Eq. \eqref{eq:lbeq_general} to vector form by replacing the single-relaxation BGK collision term
$-\frac{1}{\tau}\left( f_i\left(\mathbf{x},t\right) - f_i^{eq}\left(\mathbf{x}, t\right)\right)$ with a general collision matrix $\ve \Sigma \in \mathbb{R}^{19 \times 19}$,
\begin{equation}
\mathbb{R}^{19} \ni \bar{\mathbf{\Omega}}^{\mathrm{MRT}} \equiv \ket{\bar{\Omega}^{\mathrm{MRT}} } = -\ve \Sigma\left(\ket{f(\ve x, t)} - \ket{f^{eq}(\ve x, t)}\right),
\end{equation}
where $\ve \Sigma = \omega \ve I$, $\omega \equiv \tau^{-1}$ for the BGK collision operator. In the MRT/LBM model the collision operator relaxes the various kinetic modes individually, thus increasing the stability as the relaxation time of a mode can be adjusted to its characteristic time. This is accomplished by transforming the usual velocity-space distribution function $f_i$ to moment-space by a transformation matrix $\ve M$, $\ket{m(\ve x, t)} = \ve M \ket{f(\ve x, t)}$. 
Following \cite{el00312, 1742-5468-2010-11-P11026}, we define $m_0$ to be the fluid density, $m_2$ the energy, $\{m_3,m_5,m_7\}$ the momentum flux and $\{m_9,m_{11},m_{13},m_{14},m_{15}\}$ components of the symmetric traceless viscous stress tensor. As density and momentum flux are conserved during collision, the moments $\{m_0,m_3,m_5,m_7\}$ are identical to their equilibrium values and the remaining non-conserved equilibrium moments are written as functions of these \cite{el00312}.

With this transformation the collision operator becomes
\begin{equation}
\ket{\bar{\Omega}^{\mathrm{MRT}} } = -\ve M^{-1}\hat {\ve \Sigma}\left(\ve m^{eq}(\ve x, t) - \ve m(\ve x, t)\right),
\end{equation}
where the transformation matrix $\ve M$ is constructed such that the relaxation matrix $\hat {\ve \Sigma}=\ve M \ve \Sigma \ve M^{-1}=\text{diag}(s_0, s_1, s_2, \ldots, s_{18})$ is diagonal in moment space and specifies the relaxation time of moment $m_i$, \cite{el00312}. 

As the corresponding moments are conserved, $s_0=s_3=s_5=s_7=0$. Utilizing the values of the two relaxation time (TRT) model in \cite{Pan2006898}, we fix the remaining diagonal elements of $\hat {\ve \Sigma}$ to the values
\begin{eqnarray}
&&s_1=s_2=s_{9-15}=\omega;\\
&&s_4=s_6=s_8=s_{16-18}=8\,\frac{2-\omega}{8-\omega}.
\end{eqnarray}
%
%
We note that the bounce-back method for handling solid boundary conditions, described in Section \ref{sec:solid_bc}, is still applicable in the MRT model.

In Fig. \ref{fig:find_perm} we present the results of our MRT-LBM simulations on the porous sample. The relation between the flow rate and the pressure difference is clearly linear, as expected from Darcy's law. All experiments were performed with the Reynolds numbers on the order of unity.

\section{Conclusions} 
\label{sec:conc}
The developments of lattice Boltzmann schemes on unstructured grids are highly promising since the fact that the velocity and space discretizations are independent can be exploited to enhance the geometric flexibility and accuracy when simulating complex flows. As we have proven in this paper, the unstructured LBM is capable of accurately simulating flows in complex, three-dimensional domains (such as pore spaces in porous rocks) at low Reynolds number using significantly fewer elements than the regular grid based approaches, such as \cite{ramstad.2010}, \cite{pazdniakou.2013}. However, regular grid based LBMs are likely to remain the most widely applied variant of the method, as they are simpler to implement and analyse, as well as more readily and massively parallelizable than the unstructured LBM, \cite{Tolke:2008:TCD:1451677.1451680}, \cite{rinaldi.2012}, \cite{DBLP:journals/cphysics/JanuszewskiK14}.

Recent works on the lattice Boltzmann models suggest several strategies to further improve on the unstructured lattice Boltzmann method. \cite{Patil20095262} present an alternative approach to solving the lattice Boltzmann equation on 2D unstructured meshes. Instead of using vertex-centred finite volume method, they choose the elements (triangles) as their control volumes, which is beneficial for the implementation and the performance, as it greatly simplifies the structure of the streaming and collision matrices, as well as the solid boundary conditions. Furthermore, using a total-variation diminishing limiter allows them to increase the stability of the method and to reduce the effects of numerical diffusion.

Of particular interest is the use of Hermite multi-speed models based on the general characteristic-based algorithm for off-lattice Boltzmann simulations, \cite{Karlinoff}, which led to full freedom in the selection of the velocity model, independently from the spatial and temporal discretizations. Using this scheme, the simulations of a 2D Taylor-Green vortex flow were run up to $\mathrm{Re}=5000$ with time step size $\delta t=500\tau$, which is a clear evidence that the time step restriction was overcome. As indicated by \cite{Karlinoff}, this method could be further improved by incorporating lattice Boltzmann $H$-theorem. Considering that, the entropic lattice Boltzmann method has emerged as a robust tool for simulations of high Reynolds number flows, see e.g.: \cite{PhysRevE.75.036712}, \cite{FLM:7843223}. With the addition of novel boundary conditions, \cite{Chikatamarla20131925} developed a robust method for sub-grid simulations of wall bounded turbulent flows flows without further modelling. The latter combined together with the geometric flexibility provided by off-lattice schemes could be useful to shed light on the interplay between surface geometry (roughness) and turbulent structures in realistic high Reynolds number flows in various engineering applications.   
\section*{Acknowledgement}
We thank S. Pedersen, K. N. Dalby and D. Müter, H. Suhonen for their help with the experimental work at beamline ID22 at  the European Synchrotron Research Facility and Diwaker Jha for help with preparing the 3D mesh. Funding was provided through the grant \emph{Earth Patterns} from the Villum Foundation and by the Danish Advanced Technology Foundation and Maersk Oil and Gas A/S through the P3 project. The Danish National Research Council (via Danscatt) provided support for the experimental work.

\appendix

\section{Remarks on notation}
In this chapter we give an overview of the mathematical notation used throughout the document. We refer to scalar variables using italicized, lower case characters, such as $f_i$, $u_\alpha$, $t$ etc.; and to three-dimensional vectors using bold-face, lower case characters, e.g.: $\mathbf{u}$, $\mathbf{c}_i$, $\mathbf{r}^{jk}$. Tensors and matrices are represented with bold-face, upper case characters: $\mathbf{D}$, $\mathbf{\Sigma}$, etc. and their scalar entries are italicized, e.g.: $D_{\alpha\beta}$, $C^{jk}$. The dot product between two vectors $\mathbf{a}, \mathbf{b} \in \mathbb{R}^3$ is denoted as $\mathbf{a} \cdot \mathbf{b}$, and the tensor (outer) product of these vectors is referred to as $\mathbf{a}\mathbf{b}$, for brevity. We denote tensor contraction using ``$:$" symbol, for example
\begin{equation}
\mathbf{r}\mathbf{r} : \nabla\nabla f
\end{equation}
refers to the contraction of the outer product of vector $\mathbf{r}$ with itself, and the Hessian tensor of a scalar function $f$, i.e.
\begin{equation}
\mathbf{r}\mathbf{r} : \nabla\nabla f \equiv \sum_{\alpha = x,y,z}\sum_{\beta = x,y,z} r_\alpha r_\beta\, \partial_\alpha \partial_\beta f.
\end{equation}
\subsection{Indices and summation convention}
We use three types of indices in this document. Lower index $i$ is used exclusively to denote the variables related to the discrete velocity set, in our case, D3Q19. Hence, index $i$ can take values $0, 1, \ldots 18$. Other lower indices (typically $\alpha$, $\beta$, $\gamma$) refer to coordinates of vectors from $\mathbb{R}^3$ and tensors from $\mathbb{R}^{3 \times 3}$. Upper indices (usually $j$, $k$) are used to denote the values of discretized variables and refer to the sites at which the given variable is sampled, i.e. $f_i^j$ means the value of function $f_i$ taken at a site (vertex) $j$.

In several places throughout the \ref{app:appendixA} we switch from vector notation to coordinate-based notation, for the reader's convenience. As a consequence of our index convention, we abuse Einstein's notation in the following way. Repeated lower indices referring to coordinates (i.e. all lower indices except for $i$) in each product refer to the sum over all admissible values of these indices, in practice $x$, $y$, $z$; for example
\begin{equation}
r^{jk}_\alpha r^{jk}_\beta \partial_\alpha \partial_\beta f \equiv \sum_\alpha \sum_\beta r^{jk}_\alpha r^{jk}_\beta \partial_\alpha \partial_\beta f,
\end{equation}
however
\begin{equation}
\mathbf{c}_i f_i \neq \sum_i \mathbf{c}_i f_i.
\end{equation}
For all other types of sums, we explicitly use the $\sum$ symbol.

\section{Numerical analysis of the unstructured LBM} \label{app:appendixA}
For the purpose of analysing the properties of the schemes introduced in Section \ref{sec:numerics} we shall write them in the general form
\begin{equation}
f_i\left(\mathbf{v}^j, t+\delta t\right) = f_i\left(\mathbf{v}^j, t \right) - \delta t \sum_{k} S_i^{jk} f_i\left(\mathbf{v}^k,t \right) + \bar{\Omega}_i^j,\label{eq:scheme_general}
\end{equation}
where $\bar{\Omega}_i^j$ is the collision operator, defined as
\begin{equation}
{{\bar{\Omega}}_i^{j, \mathrm{FE}}} = - \frac{\delta t}{\tau}\sum_{k} C^{jk} \left( f_i\left(\mathbf{v}^k, t\right) - {f_i^{eq}} \left(\mathbf{v}^k, t\right) \right) \label{eqn:coll_Euler}
\end{equation}
for the forward Euler time integration, and
\begin{equation}
{{\bar{\Omega}}_i^{j, \mathrm{OS}}} = - \frac{\delta t}{\tau}\sum_{k} C^{jk} \left(  \tilde{f}_i\left(\mathbf{v}^k, t+\delta t\right) - {\tilde{f}_i^{eq}} \left(\mathbf{v}^k, t+\delta t\right) \right)
\end{equation}
for the operator splitting scheme, where 
\begin{equation}
\tilde{f}_i\left(\mathbf{v}^k, t+\delta t\right) = f_i\left(\mathbf{v}^k, t \right) - \delta t \sum_{k' \in \mathcal{N}^k} S_i^{kk'} f_i\left(\mathbf{v}^{k'},t \right)
\end{equation}
and $\tilde{f}_i^{eq}$ is evaluated using the values of $\tilde{f}_i$. In order to derive the Navier-Stokes equation from Eq. \eqref{eq:scheme_general} we perform the Chapman-Enskog expansion. For the sake of clarity, we shall first consider the streaming operator alone and then analyse the collision operators.

\subsection{Streaming operator}
\label{sec:streaming}
As the first step towards the Chapman-Enskog expansion, we consider the Taylor expansion of Eq. \eqref{eq:scheme_general} around $\left( \mathbf{v}^j, t \right)$, up to the second order terms. The Taylor expansion of the left-hand side of \eqref{eq:scheme_general} reads
\begin{equation}
f_i\left(\mathbf{v}^j, t+\delta t\right) = f_i^j + \delta t\, \partial_t f_i^j + \frac{1}{2}\delta t^2\,\partial_t^2 f_i^j+O\left(\delta t^3\right), \label{eq:taylor_time}
\end{equation}
where we use a shorthand notation $f_i^j \equiv f_i\left(\mathbf{v}^j, t \right)$. Similarly
\begin{equation}
f_i\left(\mathbf{v}^k,t \right) = f_i^j + \delta\mathbf{r}^{jk} \cdot \nabla f_i^j + \frac{1}{2} \mathbf{r}^{jk}\mathbf{r}^{jk} : \nabla \nabla f_i^j + O\left(\delta r^3\right), \label{eq:taylor_space}
\end{equation}
where $\mathbf{r}^{jk} = \mathbf{v}^k - \mathbf{v}^j$. By substituting \eqref{eq:taylor_time} and \eqref{eq:taylor_space} into \eqref{eq:scheme_general} and subtracting $f_i^j$ from both sides we obtain
\begin{equation}
\delta t\, \partial_t f_i^j + \frac{1}{2}\delta t^2\,\partial_t^2 f_i^j = - \delta t \sum_{k} S_i^{jk}\left(f_i^j + \mathbf{r}^{jk} \cdot \nabla f_i^j + \frac{1}{2} \mathbf{r}^{jk}\mathbf{r}^{jk} : \nabla \nabla f_i^j\right)+\bar{\Omega}_i^j. \label{eq:taylor_full}
\end{equation}
We can rewrite the sum on the right-hand side as
\begin{equation}
f_i^j \sum_k S_i^{jk} + \sum_k S_i^{jk} \mathbf{r}^{jk}\cdot\nabla f_i^j + \frac{1}{2}\sum_k S_i^{jk} \mathbf{r}^{jk}\mathbf{r}^{jk} : \nabla \nabla f_i^j. \label{eq:sum_streaming}
\end{equation}
Since $\forall_{i,j} \: \sum_k S_i^{jk} = 0$, the first term in \eqref{eq:sum_streaming} vanishes. Using the index notation, we can rewrite the remaining terms as
\begin{equation}
\sum_k S_i^{jk} r^{jk}_\alpha \partial_\alpha f_i^j + \frac{1}{2}\sum_k S_i^{jk} r^{jk}_\alpha r^{jk}_\beta \partial_\alpha \partial_\beta f_i^j =  \partial_\alpha f_i^j \sum_k S_i^{jk} r^{jk}_\alpha + \frac{\partial_\alpha \partial_\beta f_i^j}{2}
\sum_k S_i^{jk} r^{jk}_\alpha r^{jk}_\beta,
\end{equation}
where $\alpha, \beta$ denote the coordinates. The definition of $S_i^{jk}$ reads
\begin{equation}
\frac{1}{V^j} \oint_{\partial \Omega^j} \left( \mathbf{ c }_i \cdot \mathbf{ n } \right) f_i\, dS = \sum_k S_i^{jk} f_i^k,
\end{equation}
where $\Omega^j$ is the control volume associated with the node $\mathbf{v}^j$. Note that this equality holds for an arbitrary, continuous function which is linear over each element. In particular, it remains true if we replace $f_i$ with $\mathbf{r}^j(\mathbf{x}) = \mathbf{x} - \mathbf{v}^j$
\begin{equation}
\frac{1}{V^j} \oint_{\partial \Omega^j} \left( \mathbf{ c }_i \cdot \mathbf{ n } \right) \mathbf{r}^j \, dS = \sum_k S_i^{jk} \mathbf{r}^{jk}.
\end{equation} 
Hence
\begin{equation}
\sum_k S_i^{jk} r^{jk}_\alpha = \frac{1}{V^j} \oint_{\partial \Omega^j} \left( \mathbf{ c }_i \cdot \mathbf{ n } \right) r^j_\alpha \, dS = \frac{1}{V^j} \oint_{\partial \Omega^j} \left( r^j_\alpha \mathbf{ c }_i \right) \cdot \mathbf{ n } \, dS. 
\end{equation}
We can now apply the divergence theorem, which yields
\begin{equation}
\sum_k S_i^{jk} r^{jk}_\alpha = \frac{1}{V^j} \int_{\Omega^j} \nabla \cdot \left( r^j_\alpha \mathbf{ c }_i \right) d\Omega =  \frac{1}{V^j} \int_{\Omega^j} \mathbf{c}_i \cdot \nabla r^j_\alpha \, d\Omega = \frac{1}{V^j} \int_{\Omega^j} \mathbf{c}_i \cdot \mathbf{e}_\alpha d\Omega,
\end{equation}
where $\mathbf{e}_\alpha$ is the unit vector associated with coordinate $\alpha$. Clearly $\mathbf{c}_i \cdot \mathbf{e}_\alpha = c_{i\alpha}$, which is a constant. Thus we finally obtain
\begin{equation}
\sum_k S_i^{jk} r^{jk}_\alpha = \frac{c_{i\alpha}}{V^j} \int_{\Omega^j}  d\Omega = c_{i\alpha}. \label{eq:first_order_streaming}
\end{equation}
The second order term in \eqref{eq:sum_streaming} can be written as
\begin{equation}
\frac{1}{2}\partial_\alpha \partial_\beta f_i^j
\sum_k S_i^{jk} r^{jk}_\alpha r^{jk}_\beta = \frac{1}{2} {D^j_i}_{\alpha \beta} \, \partial_\alpha \partial_\beta f_i^j  \label{eq:second_order_streaming}
\end{equation}
where ${D_i^j}_{\alpha \beta} = \sum_k S_i^{jk} r^{jk}_\alpha r^{jk}_\beta$ is known as the \emph{numerical diffusion tensor}, \cite{ubertini_bella_succi2003}. By substituting \eqref{eq:first_order_streaming} and \eqref{eq:second_order_streaming} into \eqref{eq:taylor_full} we finally obtain
\begin{equation}
\delta t\, \partial_t f_i^j + \frac{1}{2}\delta t^2\,\partial_t^2 f_i^j = - \delta t\, c_{i\alpha} \, \partial_\alpha f_i^j - \frac{\delta t}{2}{D_i^j}_{\alpha \beta} \, \partial_\alpha \partial_\beta f_i^j+\bar{\Omega}_i^j. \label{eq:streaming_expanded}
\end{equation}

\subsubsection{Numerical diffusion tensor}
The definition of the numerical diffusion tensor as ${D_i^j}_{\alpha \beta} = \sum_k S_i^{jk} r^{jk}_\alpha r^{jk}_\beta$ is not very convenient for further analysis, due to dependence on $i$. In this section we will analyse it in greater detail. Once again, we will apply the definition of the streaming operator $S_i^{jk}$
\begin{equation}
\frac{1}{V^j} \oint_{\partial \Omega^j} \left( \mathbf{ c }_i \cdot \mathbf{ n } \right) \zeta^j_{\alpha\beta}(\mathbf{x})\, dS = \sum_k S_i^{jk} \zeta^{jk}_{\alpha\beta},
\end{equation}
where the function $\zeta^{j}_{\alpha\beta}$ is constructed in a way that $\zeta^{jk}_{\alpha\beta} = \zeta^{j}_{\alpha\beta}(\mathbf{v}^k) =  r_\alpha^{jk} r_\beta^{jk}$ and is linear in each element containing $\mathbf{v}^j$ 
\begin{equation}
\zeta^{j}_{\alpha\beta}(\mathbf{x}) = \sum_k r_\alpha^{jk} r_\beta^{jk} \phi^k(\mathbf{x}), 
\end{equation}
where $\phi^k(\mathbf{x})$ is the linear interpolant (or \emph{hat function}) associated with vertex $\mathbf{v}^k$ (i.e. $\phi^k(\mathbf{v}^k) = 1$, $\phi^k(\mathbf{v}^l) = 0$, $l \neq k$ and $\phi^k$ is linear over each element). Then, using the divergence theorem, we obtain
\begin{equation}
\sum_k S_i^{jk} \zeta^{jk}_{\alpha\beta} = \frac{1}{V^j} \oint_{\partial \Omega^j} \left( \zeta^j_{\alpha\beta} \mathbf{ c }_i \right) \cdot \mathbf{ n }\, dS = \frac{1}{V^j} \int_{\Omega^j} \nabla \cdot \left( \zeta^j_{\alpha\beta} \mathbf{ c }_i \right) d\Omega = \mathbf{c}_i \cdot \left(\frac{1}{V^j} \int_{\Omega^j} \nabla \zeta^j_{\alpha\beta} d\Omega \right).
\end{equation}
Notice that $\nabla \zeta_{\alpha\beta}^j = \sum_k r_\alpha^{jk} r_\beta^{jk} \nabla \phi^k$, and since $\phi^k$ is linear over each element, then $\nabla \phi^k$ and, in consequence, $\nabla \zeta_{\alpha\beta}^j$ is constant over each element. Let us denote 
\begin{equation}
\mathbf{\Delta}^j_{\alpha\beta} = \frac{1}{V^j}\int_{\Omega^j} \nabla \zeta_{\alpha\beta}^j d\Omega.
\end{equation}  
Such vector $\mathbf{\Delta}^j_{\alpha\beta}$ depends only on the geometry of the mesh. Then
\begin{equation}
{D_i^j}_{\alpha\beta} = \sum_k S_i^{jk} \zeta^{jk}_{\alpha\beta} = \mathbf{c}_i \cdot \mathbf{\Delta}^j_{\alpha\beta} = c_{i\gamma} \Delta^j_{\alpha\beta\gamma},
\end{equation}
finally allowing us to rewrite equation \eqref{eq:streaming_expanded} as
\begin{equation}
\delta t\, \partial_t f_i^j + \frac{1}{2}\delta t^2\,\partial_t^2 f_i^j = - \delta t\, c_{i\alpha} \, \partial_\alpha f_i^j - \frac{\delta t}{2}\Delta_{\alpha\beta\gamma}^j c_{i\gamma} \, \partial_\alpha \partial_\beta f_i^j+\Omega_i^j. \label{eq:streaming_final}
\end{equation}
Notice that the effects of numerical diffusion scale quadratically with the edge lengths $r$, and therefore disappear for well-resolved meshes.

\subsection{Collision operators}
\label{sec:collision}
\subsubsection{Forward Euler time integration}

 In contrast to the standard LB schemes developed on regular grids, we can see from expression \eqref{eqn:coll_Euler} that the collision operator in the present finite volume formulation is non-local, i.e. the relaxation towards equilibrium at a specific grid point is a function of the relaxation at the  neighbouring points. Since the space-time dependence of the equilibrium distribution is through the fluid quantities, namely the density and velocity, we can expect that for sufficiently smooth flows the hydrodynamic fields do not vary significantly on the scales of grid spacing and consequently non-local effects in evaluating the equilibrium distribution function should be negligible. Let us determine the value of these quantities at a specific grid point $\mathbf{v}^k$ as a function of their values at the neighbouring points. We first perform a Taylor expansion of the particle distribution function around the neighbouring grid point $\mathbf{v}^j$, which leads to
\begin{equation}
f_i(\mathbf{v}^k,t)\approx f_i(\mathbf{v}^j,t)+\mathbf{r}^{jk}\cdot\nabla f_i +\frac{1}{2}\mathbf{r}^{jk} \mathbf{r}^{jk}:\nabla\nabla f_i. \label{taylorexpansion}
\end{equation}
Inserting the above expansion into the definition of the density $\rho = \sum_i f_i$ we obtain 
\begin{equation}
\rho(\mathbf{v}^k,t)=\sum_i{f_i(\mathbf{v}^k,t)}\left(\approx \sum_i{f_i(\mathbf{v}^j,t)+\mathbf{r}^{jk}\cdot\nabla f_i +\frac{1}{2}\mathbf{r}^{jk} \mathbf{r}^{jk}:\nabla\nabla f_i}\right).  
\end{equation}
Interchanging the sum operation with the spatial derivative in the above equation yields
\begin{equation}
\rho(\mathbf{v}^k,t) \approx \rho^j\left(1+\frac{1}{\rho^j}\,\mathbf{r}^{jk}\cdot\nabla\rho^j +\frac{1}{2\rho^j}\,\mathbf{r}^{jk} \mathbf{r}^{jk}:\nabla\nabla\rho^j\right),
\label{densityexpansion} 
\end{equation}
where we denote $\rho(\mathbf{v}^j, t) \equiv \rho^j$ for brevity. Following the same procedure as for the density we obtain for the momentum
\begin{equation}
\rho(\mathbf{v}^k,t) \mathbf{u}(\mathbf{v}^k,t) = \sum_i f_i(\mathbf{v}^k, t) \mathbf{c}_i \approx \rho^j \mathbf{u}^j + (\mathbf{r}^{jk} \cdot \nabla) (\rho^j \mathbf{u}^j) + \frac{1}{2} \mathbf{r}^{jk} \mathbf{r}^{jk} : \nabla \nabla (\rho^j \mathbf{u}^j )
\end{equation}
Here, $\mathbf{u}^j \equiv \mathbf{u}(\mathbf{v}^j, t)$. Now, by approximating
\begin{equation}
\frac{1}{\rho(\mathbf{v}^k, t)} \approx \frac{1}{\rho^j}\left( 1 - \frac{1}{\rho^j}\,\mathbf{r}^{jk}\cdot\nabla\rho^j -\frac{1}{2\rho^j}\,\mathbf{r}^{jk} \mathbf{r}^{jk}:\nabla\nabla\rho^j \right)
\end{equation}
using the first order Taylor expansion, and by omitting the products of the derivatives of density and momentum (see the discussion below), we obtain the velocity
\begin{equation}
\mathbf{u}(\mathbf{v}^k,t)\approx \mathbf{u}^j\left(1-\frac{1}{\rho^j}\mathbf{r}^{jk}\cdot\nabla\rho^j-\frac{1}{2\rho^j}\mathbf{r}^{jk} \mathbf{r}^{jk}:\nabla\nabla\rho^j\right)+\frac{1}{\rho^j}(\mathbf{r}^{jk}\cdot\nabla)(\rho^j \mathbf{u}^j) +\frac{1}{ 2 \rho^j}\mathbf{r}^{jk} \mathbf{r}^{jk}:\nabla\nabla (\rho^j \mathbf{u}^j) \label{velexpansion} 
\end{equation}
For the sake of the simplicity of the notation let us write the above equations for the density and velocity as       
\begin{equation}
\rho(\mathbf{v}^k,t)\approx \rho^j+\Delta\rho^k,\label{densdelta}
\end{equation}
and 
\begin{equation}
\mathbf{u}(\mathbf{v}^k,t)\approx \mathbf{u}^j+\Delta \mathbf{u}^k,
\label{veldelta} 
\end{equation}
respectively. Substitution of Eqs. \eqref{densdelta} and \eqref{veldelta} into the equilibrium distribution function yields   
\begin{equation}
\begin{split}
f_i^{eq}(\mathbf{v}^k,t) \approx & w_i \left(\rho^j+\Delta\rho^k\right) \left(1+\frac{\mathbf{c}_i\cdot \mathbf{u}^j}{c_s^2}+\frac{(\mathbf{c}_i\cdot \mathbf{u}^j)^2}{2 c_s^4}-\frac{(\mathbf{u}^j)^2}{ 2 c_s^2}\right)
\\ 
+ & w_i \left(\rho^j+\Delta\rho^k\right)\left(\frac{\mathbf{c}_i\cdot \Delta \mathbf{u}^k}{c_s^2}+ \frac{(\mathbf{c}_i\cdot \Delta \mathbf{u}^k)^2                +2 \mathbf{c}_i\cdot \mathbf{u}^j (\mathbf{c}_i\cdot \Delta \mathbf{u}^k) }{2 c_s^4}-\frac{(\Delta \mathbf{u}^k)^2}{2 c_s^2}-\frac{2 \mathbf{u}^j\cdot\Delta \mathbf{u}^k}{2 c_s^2} \right)
\end{split}
\label{equilibriumsplitted}
\end{equation}
By neglecting all the quadratic terms that contain spatial derivatives of both the density and momentum, contained in $\Delta\rho^k$ and $\Delta \mathbf{u}^k$, we obtain that the equilibrium distribution can be written to leading order as\footnote{ The quadratic terms containing the spatial derivatives of both the density and the momentum are vanishingly small for well resolved meshes, as well as for flows at low Mach and Reynolds numbers. }
\begin{equation}
f_i^{eq}(\mathbf{v}^k,t) \approx f_i^{eq}(\mathbf{v}^j,t)\left(1+\frac{\Delta\rho^k}{\rho^j}\right)
+ w_i\rho^j\left(\frac{\mathbf{c}_i\cdot \Delta \mathbf{u}^k}{c_s^2}+ 
\frac{\mathbf{c}_i\cdot \mathbf{u}^j (\mathbf{c}_i\cdot \Delta \mathbf{u}^k)}{c_s^4}-\frac{\mathbf{u}^j\cdot \Delta \mathbf{u}^k}{c_s^2}\right)
\label{equilibriumsimplified}
\end{equation}
     
Further simplifications can be made by taking into account the explicit expressions for $\Delta\rho^k$ and $\Delta \mathbf{u}^k$. The key point is that the linear terms in the Taylor expansion are of the form $\mathbf{r}^{jk}\cdot\nabla$ and, as discussed in Section \ref{sec:meshing}, the sum $\sum_{k}C^{jk} \mathbf{r}^{jk}$ is very close to zero when $\mathbf{v}^j$ lies at the geometrical center of the control volume. Consequently, the only non-vanishing contribution to the sum over the control volume will be given by the second order terms in the Taylor expansions of $\Delta\rho^k$ and $\Delta u^k$. These considerations lead to the following relation
\begin{equation}
\begin{split}
& \sum_{k}{C^{jk} f_i^{eq}(\mathbf{v}^k,t)} \approx f_i^{eq}(\mathbf{v}^j,t)\left(1+\sum_{k}C^{jk}\frac{\mathbf{r}^{jk} \mathbf{r}^{jk}:\nabla\nabla\rho^j }{2 \rho^j}\right) 
\\
& - w_i\rho^j \left(1+\frac{\mathbf{c}_i\cdot \mathbf{u}^j}{c_s^2}\right) \left(\frac{\mathbf{c}_i\cdot \mathbf{u}^j}{c_s^2}\sum_{k}{C^{jk}\frac{1}{2 \rho^j}\mathbf{r}^{jk} \mathbf{r}^{jk}:\nabla\nabla \rho^j}+\frac{\mathbf{c}_i}{c_s^2}\cdot\sum_{k}{C^{jk}\frac{1}{2 \rho^j }\mathbf{r}^{jk} \mathbf{r}^{jk}:\nabla\nabla (\rho^j \mathbf{u}^j)}\right)
 \\
& -\frac{w_i\rho^j}{c_s^2}\mathbf{u}^j \cdot\left(\mathbf{u}^j\sum_{k}{C^{jk}\frac{1}{2 \rho^j }\mathbf{r}^{jk} \mathbf{r}^{jk}:\nabla\nabla \rho^j}+\sum_{k}{C^{jk}\frac{1}{2 \rho^j }\mathbf{r}^{jk} \mathbf{r}^{jk}:\nabla\nabla (\rho^j \mathbf{u}^j)}\right),
 \end{split}
 \label{eqsum}
 \end{equation}
where we have used the sum rule $\sum_{k}C^{jk} = 1$. We can see that the terms containing the sums of the second order terms of the Taylor expansion times the collision matrix are vanishingly small for well resolved meshes and flows at low Mach and Reynolds numbers, as is in our case. We therefore can safely neglect those terms and obtain
\begin{equation}
\sum_{k}{C^{jk} f_i^{eq}(\mathbf{v}^k,t)} \approx f_i^{eq}(\mathbf{v}^j,t).
\end{equation}

With this result we can further analyse the collision operator by inserting the Taylor expansion, Eq. \eqref{taylorexpansion}, into the expression for the collision Eq. \eqref{eqn:coll_Euler}, which yields
\begin{equation}
\bar{\Omega}^{j,\mathrm{FE}}_i = - \frac{\delta t}{\tau}\sum_{k} C^{jk} \left[ f_i(\mathbf{v}^j,t)+\mathbf{r}^{jk}\cdot\nabla f_i +\frac{1}{2}\mathbf{r}^{jk} \mathbf{r}^{jk}:\nabla\nabla f_i - {f_i^{eq}} \left(\mathbf{v}^j, t\right) \right]
\end{equation}
Using the relations  $\sum_{k} C^{jk}=1$ and $\sum_{k} C^{jk} \mathbf{r}^{jk}=\mathbf{0}$ we arrive to the following approximation for the collision operator 
\begin{equation}
\bar{\Omega}^{j,\mathrm{FE}}_i = - \frac{\delta t}{\tau}\left( f_i(\mathbf{v}^j,t)-f_i^{eq} \left(\mathbf{v}^j, t\right)\right) - \frac{\delta t}{\tau} \mathbf{D}^j:\nabla\nabla f_i,
\label{colsimplified}
\end{equation}
where the tensor $\mathbf{D}^j$ is defined as 
\begin{equation}
\mathbf{D}^j = \frac{1}{2}\sum_{k} C^{jk} \mathbf{r}^{jk} \mathbf{r}^{jk}. 
\end{equation}
We see that in Eq. \eqref{colsimplified} the first term is the standard BGK relaxation and the second one introduces numerical viscosity as we shall demonstrate below by doing a multi-scale analysis. Therefore, the numerical viscosity effects due to collision are of second order in $r$ for the forward Euler integration. 

We will now include the expression for the collision term $\bar{\Omega}_i^j$ for the forward Euler time-stepping scheme. Using \eqref{eq:streaming_final} and \eqref{colsimplified} the evolution equation yields
\begin{equation}
\delta t \,\partial_t f_i^j + \frac{\delta t^2}{2}\partial_t^2f_i^j = - \delta t\, c_{i\alpha}\partial_\alpha f_i^j - \frac{\delta t}{2}{\Delta}_{\alpha\beta\gamma}c_{i\gamma}\partial_\alpha\partial_\beta f_i^j - \frac{\delta t}{\tau}((f_i^j-f_i^{j, eq}) + D^j_{\alpha\beta}\partial_\alpha\partial_\beta f_i^j). \label{eqn:FE_update}
\end{equation}
We will analyse this equation in the remainder of this section. We analyse it locally, hence we drop the $j$-index. We start by multiplying both sides by ${c_{i\alpha}}$ and summing over all $i = 0,1,\ldots,18$, which yields
\begin{equation}
\partial_t (\rho u_\alpha)+ \frac{\delta t}{2}\partial_t^2 (\rho u_\alpha) = -\partial_\beta \Pi_{\alpha\beta}, \label{eqn:fe_momeq_simple}
\end{equation}
where we have introduced the momentum flux tensor $\Pi_{\alpha\beta} = \sum_i c_{i\alpha}c_{i\beta}f_i$. For small deviation around equilibrium we can write $f_i=f_i^{eq}+f_i^{neq}$, which yields
\begin{equation}
\Pi_{\alpha\beta} = \rho u_\alpha u_\beta + \rho c_s^2 + \Pi_{\alpha\beta}^{neq}
\end{equation}
and substituting this into \eqref{eqn:fe_momeq_simple} yields
\begin{equation}
\rho \partial_t u_\alpha + \rho u_\alpha \partial_\alpha u_\beta = -\partial_\alpha \rho c_s^2 - \frac{\delta t}{2}\partial_t^2\rho u_\alpha - \partial_\beta \Pi_{\alpha\beta}^{neq}.
\end{equation}
This allows us to see that the viscous stresses are contained in the term $\sum_{i}c_{i\alpha}c_{i \beta}f_i^{neq}$. By means of the Chapman-Enskog procedure we can express the viscous stress tensor in the hydrodynamic limit as a function of the fluid quantities and therefore determine the fluid viscosity. 
\subsubsection{Chapman-Enskog expansion}
Firstly, we introduce a multi-scale expansion of the distribution function around equilibrium in the small Knudsen number ($\epsilon$) limit
\begin{equation}
f_i = f_i^{(0)} + \epsilon f_i^{(1)} + \epsilon^2 f_i^{(2)} + O(\epsilon^3).
\end{equation}
Similarly, we expand the time derivation operator, separating the time scale into fast (convective) $t^{(1)}$ and slow (diffusive) $t^{(2)}$ phenomena
\begin{equation}
\partial_t = \epsilon \, \partial_{t}^{(1)} + \epsilon^2 \partial_{t}^{(2)} + O\left(\epsilon^3\right)
\end{equation}
while the spatial derivative expansion reads $\nabla = \epsilon \nabla^{(1)}$.

Dividing all terms in \eqref{eqn:FE_update} by $\delta t$ and expanding in $\eps$ gives us the following equations in the first two orders of $\eps$
\begin{eqnarray}
\eps   \,\, &:& \,\, \partial_t^{(1)} f_i^{(0)} = -c_{i\alpha}\, \partial_\alpha^{(1)} f_i^{(0)} - \frac{1}{\tau}\f{1} \label{eqn:0fe_1},  \\
\eps^2 \,\, &:& \,\, \partial_t^{(1)}  f_i^{(1)} + \partial_t^{(2)} f_i^{(0)} + \frac{\delta t}{2} {\partial_t^2}^{(1)} f_i^{(0)} = -c_{i\alpha}\, \partial_\alpha^{(1)} f_i^{(1)} - \left(\frac{{\Delta}_{\alpha\beta\gamma}}{2} c_{i\gamma} + \frac{D_{\alpha\beta}}{\tau}\right) \partial_\alpha^{(1)} \partial_\beta^{(1)} f_i^{(0)} -\frac{1}{\tau}\f{2} \label{eqn:0fe_2}
\end{eqnarray}
The zeroth velocity moments of \eqref{eqn:0fe_1} and \eqref{eqn:0fe_2} are given by 
\begin{eqnarray}
\eps   \,\, &:& \,\, \partial_t^{(1)} \rho = -\partial^{(1)}_\alpha (\rho u_\alpha),  \\
\eps^2 \,\, &:& \,\, \frac{\delta t}{2} {\partial_t^2}^{(1)}\rho + \partial_t^{(2)} \rho = -\left(\frac{{\Delta}_{pqr}c_{ir}}{2} + \frac{D_{pq}}{\tau}\right) \partial_p^{(1)}\partial_q^{(1)} \rho.
\end{eqnarray}
The first velocity moments of \eqref{eqn:0fe_1} and \eqref{eqn:0fe_2} are given by 
\begin{eqnarray}
\eps   \,\, &:& \,\, \partial_t^{(1)} (\rho u_\alpha) = -\partial^{(1)}_\beta \Pi_{\alpha\beta}^{(0)} \label{eqn:1fe_1},  \\
\eps^2 \,\, &:& \,\, \frac{\delta t}{2} {\partial_t^2}^{(1)}(\rho u_\alpha) + \partial_t^{(2)} (\rho u_\alpha) = -\partial^{(1)}_\beta \Pi_{\alpha\beta}^{(1)} -\left(\frac{{\Delta}_{pqr}c_{ir}}{2} + \frac{D_{pq}}{\tau}\right)\partial_p^{(1)}\partial_q^{(1)} (\rho u_\alpha) .
\end{eqnarray}
By neglecting the effects of the spatial discretization, we obtain from the zeroth and first velocity moments, respectively,
\begin{eqnarray}
\partial_t \rho + \partial_\alpha (\rho u_\alpha) &=& -\frac{\delta t}{2}\partial_t^2\rho, \\
\partial_t (\rho u_\alpha) + \partial_\beta \Pi_{\alpha\beta} &=& -\frac{\delta t}{2}\partial_t^2 (\rho u_\alpha).
\end{eqnarray}
Therefore we can see that mass and momentum conservation are satisfied with an error on the order of the time step. Furthermore, we remind that we have neglected all the terms of second order in the grid spacing so indicating that the errors in this ULBE scheme are linear in $\delta t$ and quadratic in $r$.

The physical viscous contribution to the hydrodynamics is governed by $\Pi_{\alpha\beta}^{(1)}$, so we will limit our analysis to this $O(\eps^1)$-term only. Our expression for $\f{1}$ is determined by \eqref{eqn:0fe_1}, which is the same expression as in regular grids. By neglecting the non-linear velocity components in the low Mach number limit, we find
\begin{equation}
\f{1} \simeq -\frac{\tau w_i}{c_s^2}(c_{i\alpha}c_{i\beta}\rho\,\partial_\alpha^{(1)} u_\beta - c_s^2\delta_{\alpha\beta}\rho\, \partial_\alpha^{(1)} u_\beta),
\end{equation}
which yields
\begin{equation}
\eps \Pi_{\alpha\beta}^{(1)} = -\rho c_s^2\tau (\partial_\alpha u_\beta + \partial_\beta u_\alpha), \label{eqn:pi1}
\end{equation}
proportional to the strain tensor. Taking the divergence of \eqref{eqn:pi1} and utilizing the assumption of incompressibility results in
\begin{equation} 
-\partial_\beta \left(\eps \Pi_{\alpha\beta}^{(1)}\right)= \rho c_s^2\tau\, \partial_\beta \partial_\beta u_\alpha, 
\end{equation}
from which it follows that the kinematic viscosity in the forward Euler scheme $\nu^{\mathrm{FE}}$ is
\begin{equation}
\nu^{\mathrm{FE}} = c_s^2\tau. \label{eqn:vis_FE}
\end{equation}

\subsubsection{Viscous stresses for the operator splitting}

 Our theoretical derivations in the above section could assessed the lack of numerical diffusivity that was observed in \cite{ubertini_bella_succi2003}. Let us investigate how the expression for the viscosity changes for the operator splitting time integration. The collision operator now reads
\begin{equation}
\bar{\Omega}^{j,\mathrm{OS}}_i = - \frac{\delta t}{\tau}\sum_{k} C^{jk} \kant[\bigg]{  \tilde{f}_i^k(t+\delta t) - \tilde{f}_{i}^{k,eq} (t+\delta t)},
\end{equation}
where
\begin{equation}
\tilde{f}_i^k(t+\delta t) = f_i^k(t ) - \delta t \sum_{k'} S_i^{kk'} f_i^{k'}(t)
\end{equation}
and $\tilde{f}_i^{eq}$ is evaluated using the values of $\tilde{f}_i$. As we have shown in the previous section, we can rewrite the definition of $\tilde f_i$ as
\begin{equation}
\tilde{f}_i^k\left(t+\delta t\right) = f_i^k\left(t \right) - \delta t \left[ c_{i\alpha} \partial_\alpha f_i^k  + \frac{1}{2} \Delta_{\alpha\beta\gamma}^k c_{i\gamma}\partial_\alpha \partial_\beta f_i^k \right],
\end{equation}
to the second order of accuracy. In order to simplify analysis, we assume that the numerical diffusion term is negligible
\begin{equation}
\tilde{f}_i^k(t+\delta t) \approx f_i^k(t ) - \delta t c_{i\alpha} \partial_\alpha f_i^k = f_i^k\left(t \right) - \delta t \, \mathbf{c}_i \cdot \nabla f_i^k = f_i^k\left(t\right) - \delta t\, \nabla \cdot \left( f_i^k \,\mathbf{c}_i\right).
\end{equation}
Now
\begin{equation}
\tilde{f}_i^{k, eq}\left(t+\delta t\right) = w_i \tilde{\rho}^k \left[ 1+ \frac{\mathbf{c}_i \cdot \tilde{\mathbf{u}}^k}{c_s^2} + \frac{\left(\mathbf{c}_i \cdot \tilde{\mathbf{u}}^k\right)^2}{2c_s^4} - \frac{\left(\tilde{u}^k\right)^2}{2c_s^2}\right], \label{eq:split_equil}
\end{equation}
where
\begin{equation}
\tilde{\rho}^k = \sum_i \tilde{f}_i^k(t+\delta t) \approx \sum_i \left[ f_i^k - \delta t\, \nabla \cdot \left( f_i^k \mathbf{c}_i \right) \right] = \rho^k - \delta t\, \nabla \cdot \left(\rho^k \mathbf{u}^k \right) = \rho^k - \delta \rho^k. \label{eq:density_sub}
\end{equation}
where $\delta \rho^k = \delta t\, \nabla\cdot \left(\rho^k \mathbf{u}^k \right)$, and
\begin{equation}
\tilde{\rho}^k \tilde{\mathbf{u}}^k = \sum_i \mathbf{c}_i \tilde{f}_i^k(t+\delta t) = \sum_i \mathbf{c}_i \left[ f_i^k - \delta t\, \nabla \cdot \left( f_i^k \mathbf{c}_i \right) \right] = \rho^k\mathbf{u}^k - \delta t\, \nabla \cdot \mathbf{\Pi}^k. \label{eq:momentum_sub}
\end{equation}
We can now rewrite Eq. \eqref{eq:split_equil} as
\begin{equation}
\tilde{f}_i^{k, eq}\left(t+\delta t\right) = w_i \left[ \tilde{\rho}^k + \frac{\mathbf{c}_i \cdot \left( \tilde{\rho}^k \tilde{\mathbf{u}}^k \right)}{c_s^2} + \frac{\left( \mathbf{c}_i \cdot \left(\tilde{\rho}^k \tilde{\mathbf{u}}^k\right)\right)^2}{2\tilde{\rho}^k c_s^4} - \frac{\left( \tilde{\rho}^k \tilde{\mathbf{u}}^k\right)^2}{2\tilde{\rho}^k c_s^2}\right],
\end{equation}
and, by substituting Eqs. \eqref{eq:density_sub}, \eqref{eq:momentum_sub}, and approximating
\begin{equation}
\frac{1}{\tilde{\rho}^k} = \frac{1}{\rho^k - \delta \rho^k} \approx \frac{1}{\rho^k} + \frac{\delta \rho^k}{\left(\rho^k\right)^2}
\end{equation}
we finally obtain
\begin{equation}
\tilde{f}_i^{k, eq}\left(t+\delta t\right) = f_i^{k, eq}(t) - \frac{\delta \rho^k}{\rho^k} f_i^{k, eq}(t) - w_i\, \delta t\, \left(\nabla \cdot \mathbf{\Pi}^k\right)\cdot\frac{\mathbf{c}_i}{c_s^2} + O\left(\mathrm{Ma}^3\right),
\end{equation} 
which can be rewritten as
\begin{equation}
\tilde{f}_i^{k,eq}\left(t + \delta t\right)\approx f_i^{k,eq}\left( t \right)\left(1-\delta t\,\frac{\partial_{\alpha}(\rho^k u^k_\alpha)}{\rho^k}\right)-\frac{\delta t\,w_i}{c_s^2}\, c_{i\gamma}\,\partial_{\alpha}(\Pi^k_{\gamma\alpha})
\label{eqdt}  
\end{equation}
In the previous section we showed that the non-local effects in the collision term are of second order in the mesh size. Since we are now concerned with the effects on the viscosity of this time discretization, for the sake of simplicity, we shall not consider any of these terms since they only depend on the spatial discretization, i.e. we approximate
\begin{equation}
\bar{\Omega}_i^{j,\mathrm{OS}} \approx -\frac{\delta t}{\tau} \left[ \tilde{f}_i^j(t+\delta t) - \tilde{f}_i^{j,eq}(t+\delta t)\right].
\end{equation}
Now the equation for the evolution of the one-particle distribution function can be written as
\begin{equation}
f_i^j(t +\delta t)= f_i^j(t)-\delta t\, c_{il}\,\partial_{l} f_i^j - \frac{\delta t}{t}(f_i^j - f_i^{j, eq}) + \frac{(\delta t)^2}{\tau}\left[ c_{il}\,\partial_{l} f_i^j-\frac{1}{\rho^j}f_i ^{eq}\,\partial_{l}(\rho^j u^j_{l})-w_i\,\partial_{l}(\Pi^j_{\gamma l})\,\frac{c_{i\gamma}}{c^2_s} \right],
\label{fullevolution}
\end{equation}
where the tensor $\Pi^j_{\gamma l}$ represent the stress tensor defined as \begin{equation}
\Pi^j_{\gamma l}=\sum_{m}c_{m\gamma}c_{m l}f^j_m
\end{equation}     
We now expand the left hand side in a Taylor series to the second order in $\delta t$, to obtain
\begin{equation}
\begin{split}
f_i^j(t)+\delta t\,\partial_{t}f^j_i+ \frac{(\delta t)^2}{2}\,\partial^2_{t}f^j_i & = f_i^j(t)-\delta t\, c_{il}\,\partial_{l} f_i^j + \frac{(\delta t)^2}{\tau}\,c_{il}\,\partial_{l} f_i^j -\frac{\delta t}{\tau}(f_i^j-f_i^{j,eq})\\ & -\frac{(\delta t)^2}{\rho\tau}f_i^{j,eq}\,\partial_{l}(\rho^j u^j_{l})-w_i\frac{(\delta t)^2}{\tau}\partial_{l}(\Pi^j_{\gamma l})\frac{c_{i\gamma}}{c^2_s}.
\end{split}
\label{evolutionwithtaylor}
\end{equation}
After simplification and dropping the $j$-index, the equation above can be written as      
\begin{equation}
\partial_{t}f_i+ \frac{\delta t}{2}\,\partial^2_{t}f_i= -c_{il}\,\partial_{l} f_i + \frac{\delta t}{\tau}c_{il}\,\partial_{l} f_i-\frac{1}{\tau}(f_i-f_i ^{eq})-\frac{\delta t}{\rho\tau}f_i ^{eq}\,\partial_{l}(\rho u_{l})-w_i\frac{\delta t}{\tau}\partial_{l}(\Pi_{\gamma l})\frac{c_{i\gamma}}{c^2_s},
\label{simplified}
\end{equation} 
Let us find the moments of the equation above. Following from the earlier definitions we can write down the following relations
\begin{equation}
\sum_{i}c_{il}\,\partial_{l} f_i = \partial_{l} \sum_{i}c_{il}f_i=\partial_{l}(\rho u_l) \label{sum1}
\end{equation} 
\begin{equation}
\sum_{i}\frac{1}{\tau}(f_i-f_i ^{eq})=0
\label{sum2}
\end{equation}
\begin{equation}
\sum_{i}\frac{\delta t}{\rho\tau}f_i ^{eq}\,\partial_{l}(\rho u_{l})=\frac{\delta t}{\tau}\,\partial_{l}(\rho u_{l})
\label{sum3}
\end{equation} 
\begin{equation}
\sum_{i}w_i\,\frac{\delta t}{c^2_s\tau}\,\partial_{l}(\Pi_{\gamma l})\,c_{i\gamma}=\frac{\delta t}{c^2_s\tau}\,\partial_{l}(\Pi_{\gamma l})\sum_{i}w_i c_{i\gamma}=0
\label{sum4}
\end{equation} 
Using the relations Eqs. (\ref{sum1}-\ref{sum4}) in Eq. \eqref{simplified} yields            
\begin{equation}
\partial_{t}\rho+ \nabla\cdot (\rho u)=- \frac{\delta t}{2}\,\partial^2_{t}\rho \label{massconservationsplitting}
\end{equation} 
In order to find the equation for the momentum conservation we multiply Eq. \eqref{simplified} by $c_{i\alpha}$ and sum over $i$. Again, we can note that the following relations hold            
\begin{equation}
\sum_{i}\frac{1}{\tau}\,c_{i\alpha}(f_i-f_i ^{eq})=0
\label{sum2m},
\end{equation}      
\begin{equation}
\sum_{i}\frac{\delta t}{\rho\tau}\,c_{i\alpha}f_i^{eq}\partial_{l}(\rho u_{l})=\frac{\delta t}{\tau}\partial_{l}(\rho u_{l})\,u_\alpha \label{sum3m},
\end{equation} 
\begin{equation}
\sum_{i}w_i\frac{\delta t}{c^2_s\tau}\partial_{l}(\Pi_{\gamma l})c_{i\gamma}=\frac{\delta t}{c^2_s\tau}\partial_{l}(\Pi_{\gamma l})\sum_{i}w_i c_{i\gamma}c_{i\alpha}=\frac{\delta t}{\tau}\partial_{l}(\Pi_{\gamma l})\delta_{\gamma\alpha}=\frac{\delta t}{\tau}\partial_{l}(\Pi_{\alpha l}).
\label{sum4m}
\end{equation} 
Then
\begin{equation}
\partial_{t}(\rho u_{\alpha})+\partial_{l}\Pi_{\alpha l} =-\frac{\delta t}{2}\,\partial^2_{t}(\rho u_{\alpha})-\frac{\delta t}{\tau}\,\partial_{l}(\rho u_{l})u_\alpha.
\label{momentumconservation}
\end{equation} 
  We can see therefore that mass and momentum conservation are satisfied with an error of the order of the time step. Furthermore, we remind that we have neglected all the terms of second order in the grid spacing so indicating that the errors in this ULBE scheme are linear in $\delta t$ and quadratic in $r$.
  
   Let us analyse more closely the momentum flux tensor $\Pi_{\alpha l}=\sum_{i}c_{i\alpha}c_{i l}f_i$. For small deviations from equilibrium we can write the $f_i s$ as   
\begin{equation}
f_i=f_i^{eq}+f_i^{neq}
\label{eqexpansion}
\end{equation} 
Inserting this into the definition of the momentum flux tensor yields
\begin{equation}
\Pi_{\alpha l}=\sum_{i}c_{i\alpha}c_{i l}(f_i^{eq}+f_i^{neq})=\rho u_{\alpha} u_l + \rho c^2_s \delta_{\alpha l}+ \sum_{i}c_{i\alpha}c_{i l}f_i^{neq}.
\label{tensorsplitting}
\end{equation}              
Substituting Eq. \eqref{tensorsplitting} into Eq. \eqref{momentumconservation} leads to
\begin{equation}
\begin{split}
u_{\alpha} \left( \partial_{t}\rho +  \partial_{l}\left(\rho u_l\right)\right)+\rho\partial_{t}u_{\alpha} +\rho u_l\partial_{l}u_{\alpha} & =-\partial_{\alpha}(\rho c^2_s)-\partial_{l}\left(\sum_{i}c_{i\alpha}c_{i l}f_i^{neq}\right)
     \\ &   -\frac{\delta t}{2}\partial^2_{t}(\rho u_{\alpha})-\frac{\delta t}{\tau}\partial_{l}(\rho u_{l})u_\alpha,
\end{split}
\label{momentumrewritten}
\end{equation} 
which, by using the mass conservation Eq. \eqref{massconservationsplitting}, can be further simplified to          
\begin{equation}
\rho\,\partial_{t}u_{\alpha} +\rho u_l\partial_{l}u_{\alpha} =-\partial_{\alpha}(\rho c^2_s)-\partial_{l}\left(\sum_{i}c_{i\alpha}c_{i l}f_i^{neq}\right) -\delta t\, \partial_{t}\rho\left(\partial_{t}u_{\alpha}-\frac{u_{\alpha}}{\tau}\right)-\frac{\delta t}{2}\rho\,\partial^2_{t}u_{\alpha}.
\label{momentumsimplified}
\end{equation}   
In the equation above the viscous stresses are contained in the term $\sum_{i}c_{i\alpha}c_{i l}f_i^{neq}$. By means of the Chapman-Enskog expansion we can express the viscous stress tensor in the hydrodynamic limit as a function of the fluid quantities and therefore determine the fluid viscosity.    
 
As usual, we can expand $f_i$ formally in terms of powers of the Knudsen numbers around the equilibrium distribution 
\begin{equation}
f_i=f_i^{(0)}+\epsilon\,f_i^{(1)} +\epsilon^2\,f_i^{(2)}.
\label{epsilonfexpansion}
\end{equation} 
For the time and spatial derivatives we can write
\begin{equation}       
\partial_t = \epsilon\,\partial_{t}^{(1)} +\epsilon^2\partial_{t}^{(2)} + O(\epsilon^3)                                        
\end{equation}
and  
\begin{equation}       
\partial_l = \epsilon\,\partial_{l}^{(1)},                                         \end{equation}
respectively. Substituting these expressions in Eq. \eqref{simplified} and collecting the terms of same power in $\epsilon$ leads to 
\begin{eqnarray} 
\epsilon^{0} &:& f_i^{(0)}=f_i^{eq}, \label{eq:f0feq}\\
\epsilon^{1} &:& \partial_{t}^{(1)}f_i^{(0)}+\frac{\tau-\delta t}{\tau}\partial_{l}^{(1)}\left(c_{il}f_i^{(0)}\right)=-\frac{1}{\tau}f_i^{(1)}-\frac{\delta t}{\rho\tau}f_i^{eq}\partial_{l}^{(1)}(\rho u_{l})-w_i\frac{\delta t}{c^2_s\tau}\partial_{l}^{(1)}(\Pi^{(0)}_{\gamma l})c_{i\gamma} \label{eq:f1exp},
\end{eqnarray}
where $ \Pi^{(0)}_{\gamma l}=\sum_{m}c_{m\gamma}c_{m l}f_m^{(0)}$. By comparing \eqref{eq:f0feq} and \eqref{epsilonfexpansion} we see that to the leading order in $\epsilon$ we have
\begin{equation}
f_i^{neq} = \epsilon f_i^{(1)},
\end{equation} 
which allows us to identify the viscous stress tensor as
\begin{equation}
\epsilon\sum_{i}c_{i\alpha}c_{i l}f_i^{(1)}.
\label{viscousstress}
\end{equation} 
From the system of equations \eqref{eq:f0feq}-\eqref{eq:f1exp} we can express $f_i^{(1)}$ in terms of $f_i^{(0)}$ (i.e. $f_i^{eq}$) in the following way  
\begin{equation} 
f_i^{(1)}=-\tau\left[\partial_{t}^{(1)}f_i^{0}+\frac{\tau-\delta t}{\tau}\partial_{l}^{(1)}\left(c_{il}f_i^{(0)}\right)+\frac{\delta t}{\rho\tau}f_i^{(0)}\partial_{l}^{(1)}(\rho u_{l})+w_i\frac{\delta t}{c^2_s\tau}\partial_{l}^{(1)}\left(\Pi^{(0)}_{\gamma l}\right)c_{i\gamma}\right] \label{f1tofo}    
\end{equation} 
Substitution of Eq. \eqref{f1tofo} into Eq. \eqref{viscousstress} yields    
\begin{equation} 
\begin{split}              
   \epsilon\Pi^{(1)}_{\alpha\kappa} & =  \epsilon\sum_{i}c_{i\alpha}c_{i\kappa}f_i^{(1)}    =-\epsilon\tau\left[\partial_{t}^{(1)}\sum_{i}c_{i\alpha}c_{i\kappa}f_i^{(0)}+\frac{\tau-\delta t}{\tau}\partial_{l}^{(1)} \left(\sum_{i}c_{i\alpha}c_{i\kappa}c_{il}f_i^{(0)}\right)\right]
    \\   &                  
     -\epsilon\tau\left[\frac{\delta t}{\rho\tau}\partial_{l}^{(1)}(\rho u_{l})\sum_{i}c_{i\alpha}c_{i\kappa}f_i^{(0)}+\frac{\delta t}{c^2_s\tau}\partial_{l}^{(1)}\Pi^{(0)}_{\gamma l}\sum_{i}w_ic_{i\alpha}c_{i\kappa}c_{i\gamma}\right] 
\end{split}     \label{viscoustensor}    
\end{equation} 
By applying the following relations fulfilled by the velocity discretization
\begin{equation}  
\sum_{i}w_i c_{i\alpha}c_{i\kappa}c_{i\gamma} = 0,
\end{equation} 
\begin{equation} 
\sum_{i}c_{i\alpha}c_{i\kappa}c_{il}f_i^{(0)}=\rho c^2_s\left(u_{\alpha}\delta_{\kappa l}+u_{\kappa}\delta_{\alpha l}+u_{l}\delta_{\alpha\kappa }\right) + O(u^{3}), 
\end{equation} 
after simplification, we obtain 
\begin{equation} 
\begin{split}              
\epsilon\Pi^{(1)}_{\alpha\kappa} & =-\epsilon\tau\left[-\frac{\tau-\delta t}{\tau}\partial_{l}^{(1)} (\rho u_{l})\left(u_{\alpha}u_{\kappa}+c^2_s\delta_{\alpha\kappa }\right) +\rho\,\partial_{t}^{(1)}(u_{\alpha}u_{\kappa})  
+\frac{\tau-\delta t}{\tau}\partial_{l}^{(1)}\left(\rho c^2_s\left(u_{\alpha}\delta_{\kappa l}+u_{\kappa}\delta_{\alpha l}\right)\right)\right] \\  &-\epsilon\tau\left[ \frac{\tau-\delta t}{\tau}c^2_s\,\partial_{l}^{(1)}\left(\rho u_{l}\right)\delta_{\alpha\kappa }\right]
\end{split}     \label{viscoussimplified1}    
\end{equation}  
Note that the terms $\left(\rho\partial_{l}^{(1)} u_{l}\right)u_{\alpha}u_{\kappa}\sim O(\M^3)$, $\left(u_{l}\partial_{l}^{(1)}\rho\right)u_{\alpha}u_{\kappa}\sim O(\M^5)$ and
$\rho u_{\kappa}\partial_{t}^{(1)}u_{\alpha}=u_{\kappa}\partial_{l}\left(\Pi^{(0)}_{\alpha l}\right)\sim O(\M^3)$ can all be neglected in correspondence with the small velocity expansion of $f_i$. After further, straightforward simplifications we end up with the following expression for $\Pi^{(1)}_{\alpha\kappa}$.
\begin{equation} 
\epsilon\Pi^{(1)}_{\alpha\kappa}=-\rho c^2_s (\tau-\delta t)\left(\partial_{\alpha} u_{\kappa}+ \partial_{\kappa} u_{\alpha}\right)
                   \label{viscoussimplified2} 
                      \end{equation}  
From the above expression we can see that the kinematic viscosity in the operator splitting scheme is equal to 
\begin{equation}
\nu^{\mathrm{OS}}=c^2_s (\tau-\delta t).
\end{equation} 
Then we see that this time discretization introduces a shift in the viscosity, reminiscent to the one in standard, regular grid based LBM schemes, in which $\nu=c^2_s\left(\tau-\frac{\delta t}{2}\right)$. Hence, with this choice of time stepping we can attain low viscosities without having to resort to prohibitively small values of $\tau$ and $dt$. Nonetheless, we can see that in order to safely neglect the errors introduced by the spatial discretization we need smooth flows on the grid spacing characteristic scales. That would mean that in order to simulate turbulent flows the mesh size must be at least of the order of the Kolmogorov scale.

\section*{References}
\bibliographystyle{elsarticle-harv} 
\bibliography{unstructlbm}

\end{document}